\documentclass[preprint]{elsarticle}

\biboptions{sort&compress}

\usepackage{hyperref}

\def\equationautorefname~#1\null{Eq.~(#1)\null}

\usepackage{amssymb}

\usepackage{enumerate}

\usepackage{caption}
\usepackage{subcaption}
\RequirePackage{multirow}
\usepackage{siunitx}
\usepackage{mathtools}

\usepackage{upgreek}

\usepackage{pifont}
\usepackage{color}

\RequirePackage{siunitx}
\DeclareSIUnit\cps{cps}
\DeclareSIUnit\gauss{G}
\DeclareSIUnit\CL{C.L.}
\journal{Astroparticle Physics}

\begin{document}


\begin{frontmatter}



\title{Muon-induced background in the KATRIN main spectrometer}


\address[a]{Technische Universit\"{a}t M\"{u}nchen, James-Franck-Str. 1, 85748 Garching, Germany}
\address[b]{Helmholtz-Institut f\"{u}r Strahlen- und Kernphysik, Rheinische Friedrich-Wilhelms Universit\"{a}t Bonn, Nussallee 14-16, 53115 Bonn, Germany}
\address[c]{Institute of Experimental Particle Physics~(ETP), Karlsruhe Institute of Technology~(KIT), Wolfgang-Gaede-Str. 1, 76131 Karlsruhe, Germany}
\address[d]{Institut f\"{u}r Physik, Johannes-Gutenberg-Universit\"{a}t Mainz, 55099 Mainz, Germany}
\address[e]{Institute for Data Processing and Electronics~(IPE), Karlsruhe Institute of Technology~(KIT), Hermann-von-Helmholtz-Platz 1, 76344 Eggenstein-Leopoldshafen, Germany}
\address[f]{Institute for Nuclear Physics~(IKP), Karlsruhe Institute of Technology~(KIT), Hermann-von-Helmholtz-Platz 1, 76344 Eggenstein-Leopoldshafen, Germany}
\address[g]{Institute for Nuclear Research of Russian Academy of Sciences, 60th October Anniversary Prospect 7a, 117312 Moscow, Russia}
\address[h]{Institute for Technical Physics~(ITeP), Karlsruhe Institute of Technology~(KIT), Hermann-von-Helmholtz-Platz 1, 76344 Eggenstein-Leopoldshafen, Germany}
\address[i]{Max-Planck-Institut f\"{u}r Kernphysik, Saupfercheckweg 1, 69117 Heidelberg, Germany}
\address[j]{Max-Planck-Institut f\"{u}r Physik, F\"{o}hringer Ring 6, 80805 M\"{u}nchen, Germany}
\address[k]{Laboratory for Nuclear Science, Massachusetts Institute of Technology, 77 Massachusetts Ave, Cambridge, MA 02139, USA}
\address[l]{Center for Experimental Nuclear Physics and Astrophysics, and Dept.~of Physics, University of Washington, Seattle, WA 98195, USA}
\address[m]{Nuclear Physics Institute of the CAS, v. v. i., CZ-250 68 \v{R}e\v{z}, Czech Republic}
\address[n]{Institut f\"{u}r Kernphysik, Westf\"{a}lische Wilhelms-Universit\"{a}t M\"{u}nster, Wilhelm-Klemm-Str. 9, 48149 M\"{u}nster, Germany}
\address[o]{Department of Physics, Faculty of Mathematics und Natural Sciences, University of Wuppertal, Gauss-Str. 20, D-42119 Wuppertal, Germany}
\address[p]{Department of Physics, Carnegie Mellon University, Pittsburgh, PA 15213, USA}
\address[q]{Universidad Complutense de Madrid, Instituto Pluridisciplinar, Paseo Juan XXIII, n\textsuperscript{\b{o}} 1, 28040 - Madrid, Spain}
\address[r]{Department of Physics and Astronomy, University of North Carolina, Chapel Hill, NC 27599, USA}
\address[s]{Triangle Universities Nuclear Laboratory, Durham, NC 27708, USA}
\address[t]{Commissariat \`{a} l'Energie Atomique et aux Energies Alternatives, Centre de Saclay, DRF/IRFU, 91191 Gif-sur-Yvette, France}
\address[u]{University of Applied Sciences~(HFD)~Fulda, Leipziger Str.~123, 36037 Fulda, Germany}
\address[v]{Department of Physics, Case Western Reserve University, Cleveland, OH 44106, USA}
\address[w]{Institute for Nuclear and Particle Astrophysics and Nuclear Science Division, Lawrence Berkeley National Laboratory, Berkeley, CA 94720, USA}
\address[x]{Institut f\"{u}r Physik, Humboldt-Universit\"{a}t zu Berlin, Newtonstr. 15, 12489 Berlin, Germany}
\address[y]{Project, Process, and Quality Management~(PPQ), Karlsruhe Institute of Technology~(KIT), Hermann-von-Helmholtz-Platz 1, 76344 Eggenstein-Leopoldshafen, Germany    }

\fntext[fn1]{Also affiliated with Oak Ridge National Laboratory, Oak Ridge, TN 37831, USA}

\author[a]{K.~Altenm\"{u}ller}
\author[b]{M.~Arenz}
\author[c]{W.-J.~Baek}
\author[d]{M.~Beck}
\author[e]{A.~Beglarian}
\author[f]{J.~Behrens}
\author[e]{T.~Bergmann}
\author[g]{A.~Berlev}
\author[h]{U.~Besserer}
\author[i]{K.~Blaum}
% ...

\newpage

\begin{abstract}

The KArlsruhe TRItium Neutrino (KATRIN) experiment aims to make a model-independent determination of the effective electron antineutrino mass with a sensitivity of 0.2 eV/c$^{2}$.
It investigates the kinematics of $\upbeta$-particles from tritium $\upbeta$-decay close to the endpoint of the energy spectrum. 
Because the KATRIN main spectrometer (MS) is located above ground, muon-induced backgrounds are of particular concern. 
Coincidence measurements with the MS and a scintillator-based muon detector system confirmed the model of secondary electron production by cosmic-ray muons inside the MS. 
Correlation measurements with the same setup showed that about \SI{12}{\%} of secondary electrons emitted from the inner surface are induced by cosmic-ray muons, with approximately one secondary electron produced for every 17 muon crossings.
However, the magnetic and electrostatic shielding of the MS is able to efficiently suppress these electrons, and we find that muons are responsible for less than \SI{17}{\%} (\SI{90}{\%} confidence level) of the overall MS background.

\end{abstract}

\begin{keyword}
	
	cosmic-ray muon backgrounds \sep electrostatic spectrometer \sep neutrino mass



\end{keyword}

\end{frontmatter}





\section{Introduction}
\label{sec:introduction}

The discovery of neutrino oscillations~\cite{Wendell2010, Aharmim2013} and the accompanying fact of neutrino mass have made the determination of the absolute neutrino mass scale an important measurement in physics. 
Investigations of the kinematics of $\upbeta$-decay provide a nearly model-independent method to determine the effective mass of electron antineutrinos. The best upper limit so far is about $\SI{2}{\electronvolt\per c\squared}$ (\SI{95}{\percent} C.L.), measured by the Mainz~\cite{Kraus2005} and the Troitsk~\cite{Lobashev2003} experiments.
Both experiments used a tritium source and a spectrometer of MAC-E filter\footnote{Magnetic Adiabatic Collimation combined with an Electrostatic filter} type \cite{Beamson1980,Lobashev1985,Picard1992}. The KArlsruhe TRItium Neutrino experiment (KATRIN) is a next-generation experiment based on the same technique, which aims to determine the effective mass of the electron antineutrino with a sensitivity of \SI{0.2}{\electronvolt\per c\squared} (\SI{90}{\percent} C.L.)~\cite{KATRIN2005}.

To achieve such a high sensitivity, it is essential to have a low background level.
As the experiment is built above ground, cosmic-ray muons could be a dominant background source\footnote{At sea level, muons are the most prevalent particle induced by cosmic-rays (ignoring neutrinos~\cite{Gaisser2016}), followed by neutrons and electrons~\cite{Bogdanova2006}. In this paper, the cosmic-ray background contribution is assumed to originate entirely from muons, although the contribution from other particles should roughly scale with the number of muons.
Thus, the effect of other cosmic-ray particles is implicitly contained in the muon rate correlation analysis discussed in \autoref{sec:correlationanalysis}.}.
The average muon flux $f$ at sea level is about 189~$\upmu$/\si{\meter\squared\per\s}~\cite{Bogdanova2006}.
The differential flux roughly follows a $\cos^2\theta$ distribution, where $\theta$ is the angle between the muon's momentum and the normal of the Earth's surface~\cite{Gaisser2016,Tanabashi2018,Formaggio2004,Chatzidakis2015}:
\begin{equation}
\frac{\text{d}f}{\text{d}\Omega}(\theta) \propto \cos^2\theta \Rightarrow
\frac{\text{d}f}{\text{d}\theta}(\theta) \propto \cos^2\theta\sin\theta,
 \label{eq:AzimuthalDistribution}
\end{equation}
where $\text{d}\Omega$ is the solid angle.
Using this distribution, a very simple \textsc{Geant4} simulation~\cite{Agostinelli2003, Allison2006, Allison2016} was performed to estimate the flux of muons through the KATRIN main spectrometer, resulting in a total rate of \num{45000}~$\upmu$/\si{\second}.
These muons produce secondary electrons as they make two crossings of the inner surface of the stainless steel vessel. 

Typical muon energies at sea-level are on the order of a \si{\giga\eV}~\cite{Tanabashi2018}, indicating that these muons primarily interact with matter via ionization~\cite{Groom2001,Bogdanov2006}.
The ionization electrons can have energies on the \si{\mega\eV} scale~\cite{PhDLeiber2014,Bogdanov2006}, but scattering and cascade processes in the material will result in most electrons emitted from the surface having energies below $\SI{30}{\electronvolt}$, with peak energies around 1--2~eV~\cite{Henke1977}.
These emitted electrons, also known as ``slow'' or ``true'' secondaries~\cite{Chuklayaev1987,Furman2002}, are accelerated by the electric field as they leave the spectrometer. 
Consequently, true secondaries that reach the detector have similar energies as the signal electrons from tritium $\upbeta$-decay and, therefore, contribute to the background.
For the Mainz neutrino experiment, muon coincidence studies indicated that secondary electrons from cosmic-ray muons contributed a significant portion of the observed background rate~\cite{Kraus2005}.

In order to investigate the muon-induced background for KATRIN, a muon detector system was installed in the spectrometer hall. 
With such an apparatus, two complementary approaches to examine the muon background are available. 
First, one can look for electron events that are coincident with events from the muon detectors.
If muons contribute to the background, a surplus of secondary electrons is expected in the time window following a signal from the muon detectors.
However, this method fails if muon-induced secondaries are trapped in the spectrometer for a significant time before being detected.
A second approach is to use the fact that the muon flux shows variations in time (on the order of hours or days) due to changes in atmospheric pressure and temperature~\cite{PhDLeiber2014}. 
Thus, one expects the background electron rate to vary in a correlated manner with the muon detector rate, if the background rate is at least partly muon-induced. 
Both of these methods (coincidence and correlation) were employed to study the muon component of the background electron rate.

\autoref{sec:apparatus} gives an introduction to the components of the KATRIN experiment, and in \autoref{sec:measurement} there is an overview of the background measurements that are relevant to this paper.
In \autoref{sec:bkgmechanism} we describe a validation of the hypothesized mechanism for muon-induced backgrounds.  
A correlation study of muon and background electron rates is then presented in \autoref{sec:correlationanalysis}, with an estimate of the remaining muon-induced background under normal KATRIN operating conditions in \autoref{sec:residualbkg}. 
Finally, in \autoref{sec:conclusion} we discuss the relevance of the muon-induced background component for KATRIN.

\section{Measurement apparatus}
\label{sec:apparatus}

\begin{figure*}
	\centering
	\includegraphics[width=1.0\textwidth]{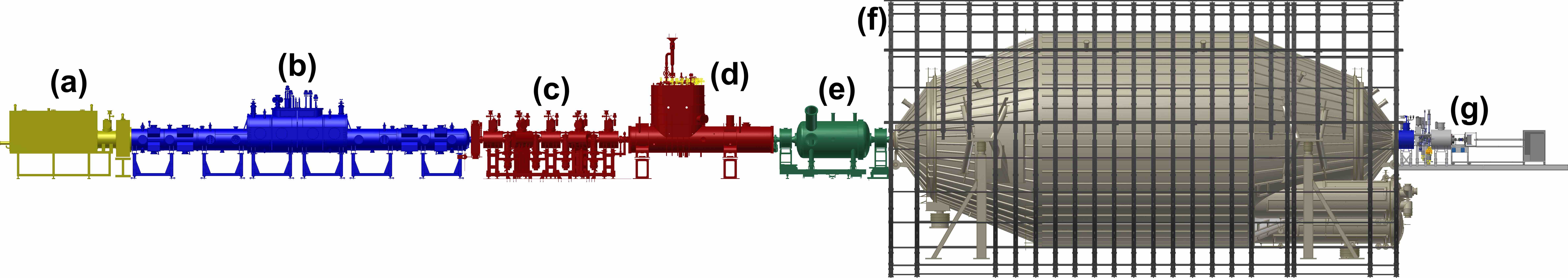}
	\caption[KATRIN experimental setup]{The KATRIN experimental setup with its primary components: (\textbf{a}) rear section; (\textbf{b}) windowless gaseous tritium source; (\textbf{c}) differential pumping section; (\textbf{d}) cryogenic pumping section; (\textbf{e}) pre-spectrometer; (\textbf{f}) main spectrometer, with air coils; (\textbf{g}) focal-plane detector.}
	\label{fig:KATRINBeamLine}
\end{figure*}

The KATRIN experiment is located at the Karlsruhe Institute of Technology, Campus North, near the city of Karlsruhe, Germany.  
The beamline (see \autoref{fig:KATRINBeamLine}) has an overall length of about \SI{70}{\m}. Molecular tritium is injected into the windowless gaseous tritium source (\textbf{b}) where it decays with an activity of 10$^{11}$~Bq, thus providing a sufficient number of $\upbeta$-decay electrons close to the endpoint energy $E_0 \approx \SI{18.6}{\kilo\electronvolt}$. 
The rear section (\textbf{a}) is responsible for monitoring the source activity and also can produce electrons for transmission studies.
The tritium is removed from the beamline in the differential pumping section (\textbf{c}) and the cryogenic pumping section (\textbf{d}) while electrons from the source are magnetically guided towards the spectrometer section. Both pre-spectrometer and main spectrometer (MS) are operated as electrostatic retarding high-pass filters of MAC-E filter type. The pre-spectrometer (\textbf{e}) is operated as a pre-filter that reduces the flux of electrons into the MS (\textbf{f}), which performs the energy analysis of the $\upbeta$-decay electrons near the endpoint ($E_0$) with an energy resolution of $\Updelta E \approx \SI{1}{\eV}$. The MS is equipped with a dual-layer wire electrode system for electrostatically shielding secondary electrons from the inner vessel surface~\cite{PhDPrall2011, PhDSchwamm2004}. The transmitted $\upbeta$-decay electrons are counted in the focal-plane detector (FPD) system (\textbf{g}) with a segmented silicon detector \cite{Amsbaugh2015}.

The MS and the FPD system are described in \autoref{subsec:MS+FPD}. Details of the muon detector system are presented in~\autoref{subsec:muonsys}.

\subsection{Main spectrometer and focal-plane detector}
\label{subsec:MS+FPD}
With a volume of about $\SI{1240}{\cubic\meter}$ and an internal surface area of about $\SI{690}{\square\meter}$, the MS is the largest component of the KATRIN experiment. 
The steel vessel has a length of about \SI{23}{\m} and a central inner diameter of \SI{9.8}{\m}~\cite{KATRIN2005}.
It works as a MAC-E filter for the energy analysis of signal $\upbeta$-particles.
Superconducting magnets at both ends of the MS generate a guiding magnetic field~\cite{Arenz2018a}. 
Signal electrons from the tritium source are guided adiabatically along the field lines towards the detector, always traveling within a flux tube delineated by the local magnetic field. 
An electrostatic retarding potential $U_0$ is applied to the MS vessel, such that only electrons with sufficient energy to overcome the resulting potential barrier reach the detector. 
The potential reaches its largest value in the vertical analyzing plane in the middle of the MS.
 For the neutrino mass measurements, $U_0$ will be varied around $\SI{-18.6}{\kilo\volt}$ in order to scan the $\upbeta$-spectrum close to this endpoint energy.

The thickness of the MS walls varies between \SI{25}{\milli\metre} and \SI{32}{\milli\metre}~\cite{KATRIN2005}.
The vessel is operated under ultra-high-vacuum conditions~\cite{MSVacuum2016} in order to minimize the energy loss from scattering of signal electrons off residual gas molecules. 
An air-coil system, consisting of 14 axial coils and two Earth's magnetic field compensation coils, is installed around the MS for the fine tuning of the magnetic field~\cite{Glueck2013, Erhard2017}. 
The polarity of each air coil can be reversed which allows a large variety of magnetic field configurations. Of particular interest for the measurements presented here are the so-called ``asymmetric configurations''~\cite{Erhard2017}, in which the magnetic field lines connect parts of the inner MS surface to the FPD (see \autoref{sec:measurement}). In this non-standard running mode, there is no flux tube connecting the entrance and exit of the MS.

\begin{figure}
  \centering
    \includegraphics[width=\columnwidth]{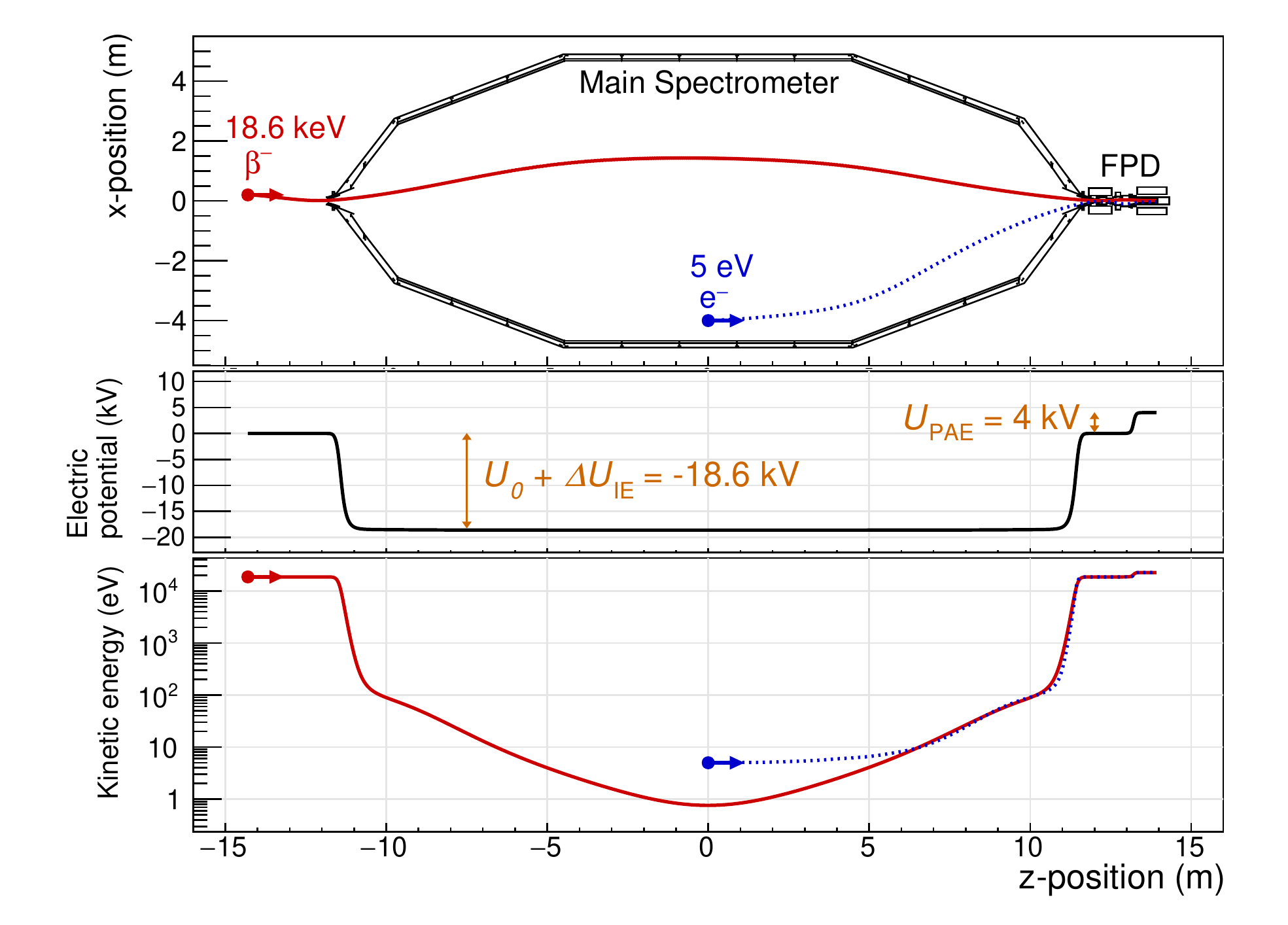}
    \caption{Schematic of electron transport inside the MS.  
    The upper plot shows two particle tracks: a through-going $\upbeta$-particle (solid red line) and a secondary electron produced inside the vessel (dotted blue line).
    The electrons spiral around the magnetic field lines as they travel, although this motion is too small to be seen in the plot.
    The middle plot shows the electric potential along the $\upbeta$-particle track, with labels indicating the important voltage contributions.
     The lower plot shows the energy of the two particles as a function of z-position.
     Due to the finite energy resolution of the FPD, the two particles cannot be distinguished.
    }
     \label{fig:PotentialSchematic}
     \end{figure}

During KATRIN operation, one possible background source comes from low-energy secondary electrons that originate from the inner MS surface.
If these electrons enter the flux tube, they can be accelerated toward the detector by the retarding potential $U_0$, in a similar way to the signal electrons.
This process is shown schematically in \autoref{fig:PotentialSchematic}.
Because the signal electrons have very low energies in most of the MS volume due to the operation of the MAC-E filter, the signal electrons cannot be distinguished from the background electrons originating from MS walls.

In the standard configuration, the magnetic field lines inside the MS are axially symmetric and approximately parallel to the walls.  
This causes charged particles (e.g. secondary electrons) emitted from the MS walls to be deflected by the Lorentz force back towards the MS surface, or, under favorable circumstances, to follow peripheral field lines outside the flux tube covered by the detector. 
Hence the magnetic guiding field provides a powerful shield against background electrons emitted from the walls.
Additional shielding is provided by an inner wire electrode (IE) system installed in two layers, close to the inner walls of the vessel~\cite{Valerius2010}. The IE system can be held at a negative potential offset $\Updelta U_{\textrm{IE}}$ of up to a few hundred volts relative to the voltage on the MS vessel, reflecting low-energy, negatively charged secondaries back toward the MS surface.
     
The FPD system~\cite{Amsbaugh2015} is situated at the downstream end of the MS. The heart of this system is the detector wafer, a silicon PIN diode whose 90-mm-diameter active area is segmented into a dartboard pattern of 148 pixels, each with an area of \SI{44}{\milli\meter\squared}. After the detector signals are amplified, a cascade of two trapezoidal filters~\cite{Jordanov1994} is applied in order to extract energy and timing information for recording via the ORCA data-acquisition software~\cite{how04}. 
An energy resolution of \SI{1.52\pm0.01}{\kilo\electronvolt} (full-width half maximum, FWHM) has been achieved for 18.6-keV electrons with this system; the FWHM timing resolution is \SI{246}{\nano\second} for a typical \num{6.4}-\si{\micro\second} shaping length~\cite{Amsbaugh2015}.
Due to an unstable preamplifier module, six detector pixels were excluded from the data analysis described in this paper.

While traveling from the MS to the FPD system, electrons must pass through a funnel-shaped post-acceleration electrode (PAE), allowing them to be accelerated by up to 10~keV.  
By increasing the energy of the electrons, one can then apply an energy cut to separate these electrons from lower-energy background electrons produced in the FPD system. 
A superconducting magnet with a \SI{3.6}{\tesla} field focuses electrons onto the detector.

\subsection{Muon detector system}
\label{subsec:muonsys}

\begin{figure*}
 \centering
  \includegraphics[width=0.9\textwidth]{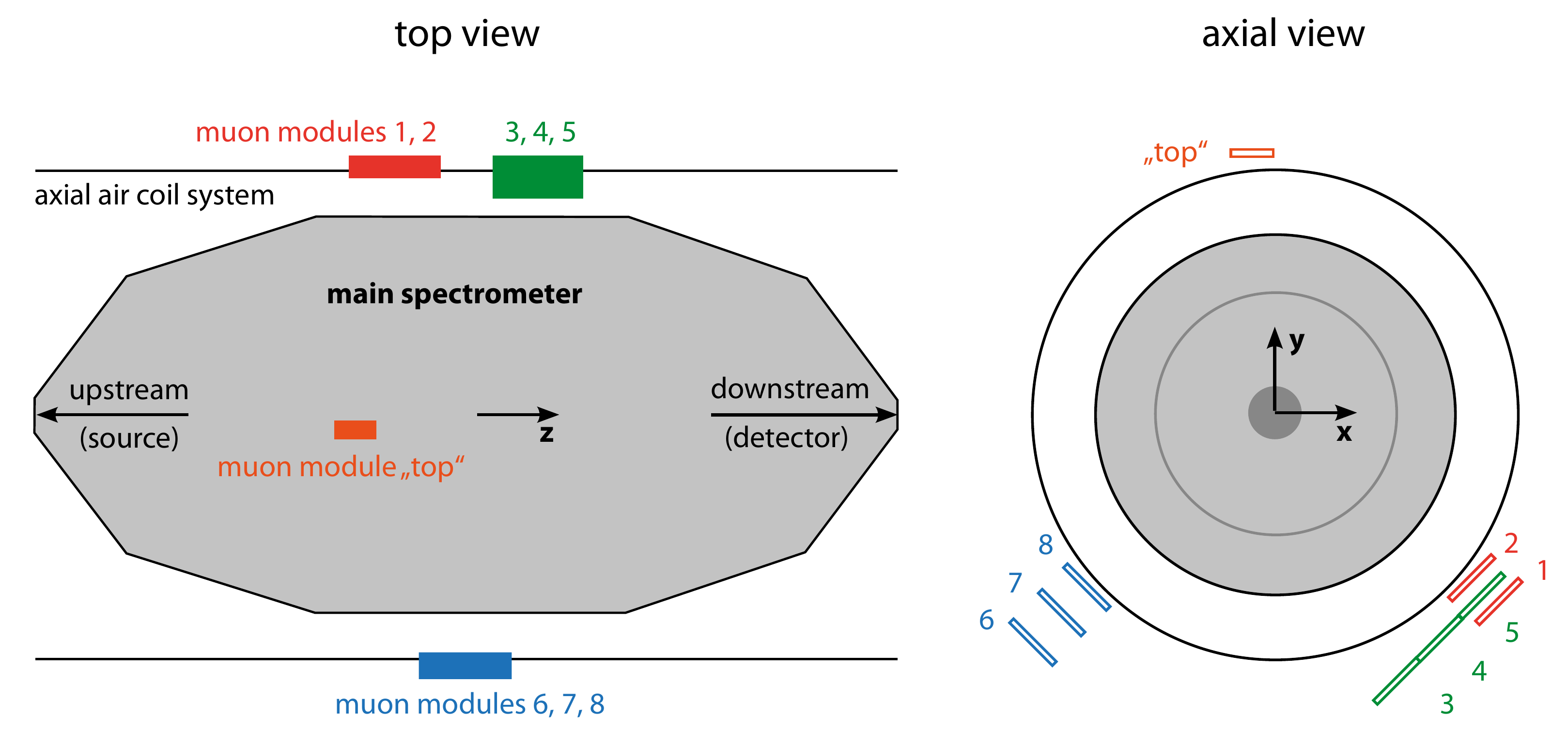}
 \caption[Muon module locations]{Location of muon modules with respect to the MS.
 Modules 6, 7, and 8 were used for the coincidence analysis.
 For the correlation study, all modules except 1 and 2 were used.
 To give a sense of scale, the large modules (1-8) have a length of \SI{3.15}{\m} (in the z-direction) and a width of \SI{0.65}{\m}.
  }
 \label{fig:MuonModuleLocations}
\end{figure*}

The muon detector system consists of eight large BICRON BC-412 plastic scintillator modules arranged in three towers around the MS vessel, as well as one smaller module above the vessel (\autoref{fig:MuonModuleLocations}). 
These modules were repurposed from the muon veto counters used in the KARMEN experiment~\cite{Drexlin1998}.
Each of the large rectangular modules has a sensitive area of $\SI{2.05}{\square\metre}$ and is equipped with four photomultipliers (PMTs) at both ends. 
The smaller module has a sensitive area of $\SI{0.3}{\square\metre}$ and only uses two PMTs located on a single side of the module.
The PMTs are wrapped in several layers of permalloy foil in order to shield them from magnetic fields near the MS. 

A muon passing through the scintillating material will induce about \num{8500} photons per \si{\MeV} of deposited energy~\cite{PhDWandkowsky2013}, and these photons are detected by the PMTs.
A dedicated ORCA DAQ system processes the signals from the PMTs and is configured to trigger on coincident events that are measured at both ends of a large scintillator module or in both PMTs for the smaller module.
Signals are collected in 50-ns time bins. 
Due to leaks in the permalloy shielding, modules 1 and 2 showed a rate dependence on the magnetic field; signals from these modules are therefore excluded from the analyses in this paper.

In order to synchronize the FPD and the muon detector systems, both DAQ systems are driven by the same precision clock. The clock provides a 10~MHz reference signal, as well as a pulse-per-second signal. Both signals are routed to the DAQ systems via optical fiber cables of equal length (50~m). The synchronization between the systems is accurate to 50~ns. An independent electronic pulser (about 0.07~Hz) is connected via BNC cables of equal length to both DAQ systems. A comparison of the timestamps of these pulser events allows the detection of possible time offsets between the two systems. 

\section{Measurements}
\label{sec:measurement}

\begin{figure*}
  \centering
      \includegraphics[width=0.49\textwidth]{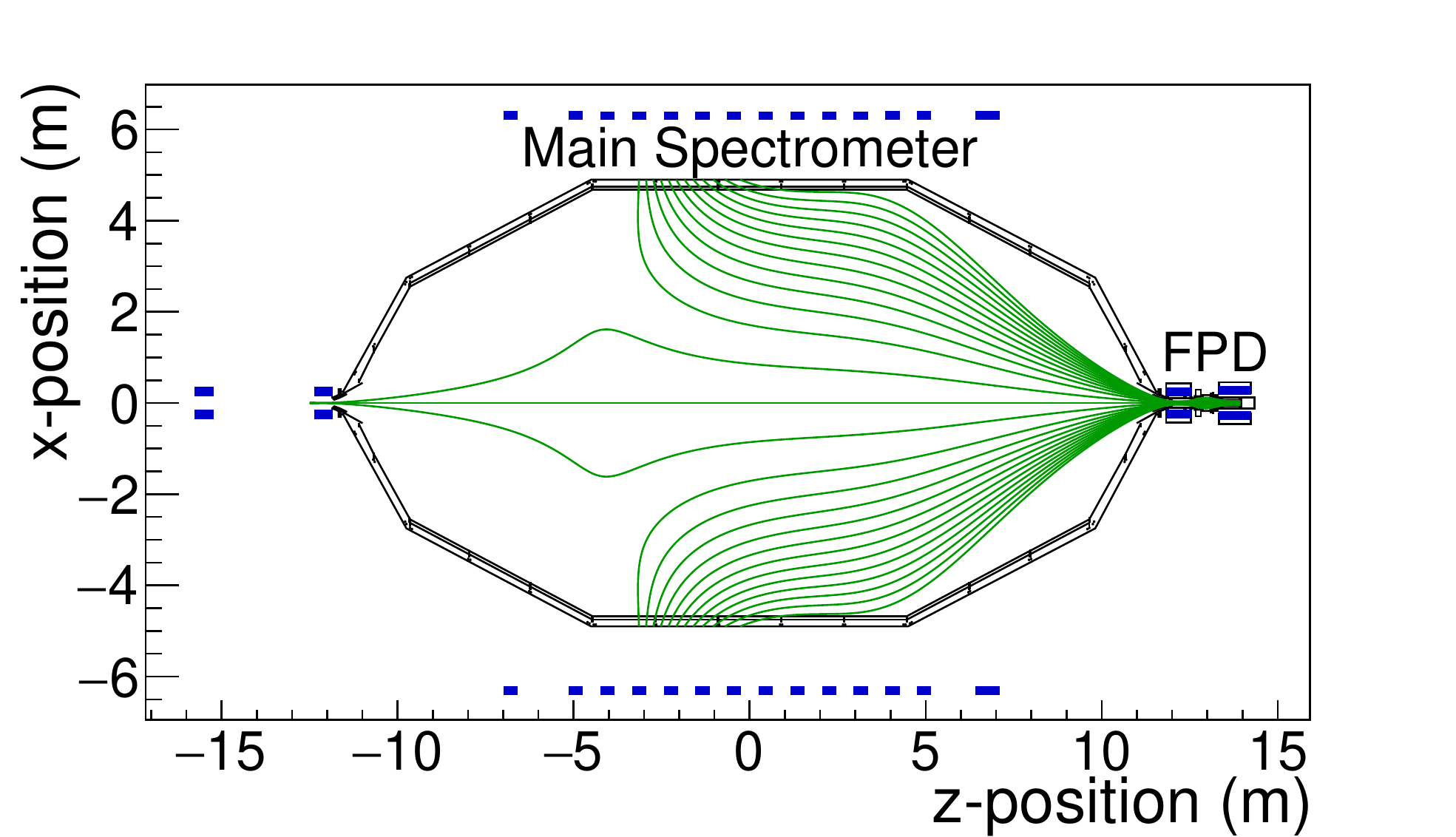}
    \includegraphics[width=0.49\textwidth]{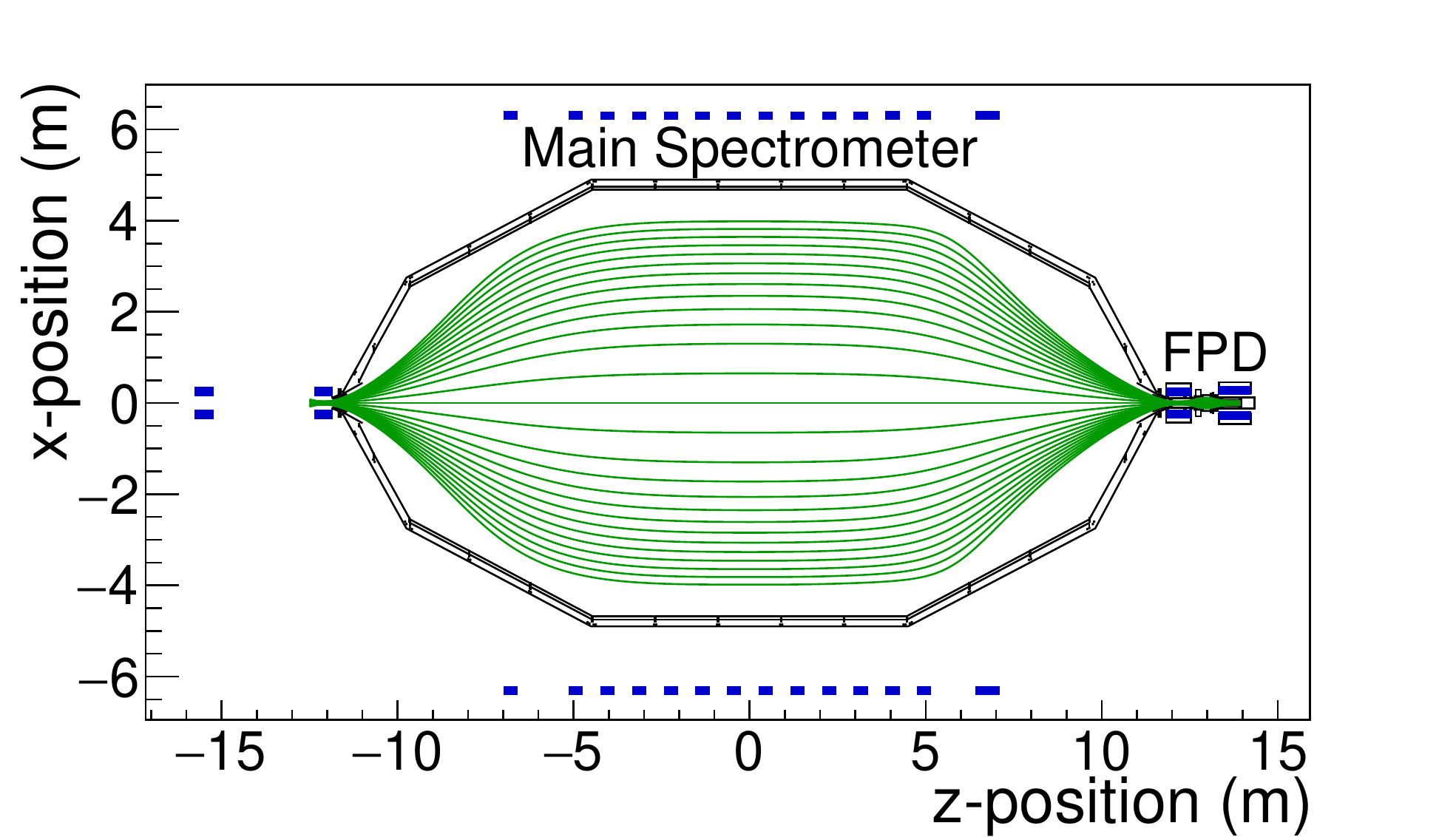}
    \caption[Magnetic field settings]{(\textit{Left}): Magnetic field lines used for setting~1 (``asymmetric'' configuration). 
    (\textit{Right}): Magnetic field lines used for settings 2 and 3 (``symmetric'' configuration). 
            The displayed field lines intersect the FPD.
    Both configurations are rotationally symmetric about the spectrometer axis, but only the ``symmetric'' setting has reflection symmetry across the $z=0$ plane.
    The blue rectangles indicate the positions of the air coils and the superconducting magnets at the MS entrance and exit.}
    \label{fig:MagneticSettings}
     \end{figure*}

The muon and FPD systems were simultaneously operated under three different electromagnetic configurations, as shown in \autoref{table:RunSettings}. 
In setting~1, an asymmetric magnetic field (\autoref{fig:MagneticSettings}, left panel) was generated where the field lines connect the surface of the MS to the active area of the FPD, maximizing the detection efficiency of electrons generated on or near the vessel walls.
In contrast, settings 2 and 3 utilized a symmetric magnetic field (\autoref{fig:MagneticSettings}, right panel) which provides magnetic shielding. 
These latter two settings are similar to the configuration to be used during KATRIN neutrino mass measurements and thus provided a more realistic background scenario.
Settings 2 and 3 differed in terms of the electrostatic shielding applied by the IE system.
Due to technical limitations of the available power supply, $\Updelta U_{\textrm{IE}} = 0$ was not possible during the measurements; the smallest stable voltage, $\Updelta U_{\textrm{IE}} = \SI{-5}{\V}$, was used instead.

\begin{table*}
\centering
  \begin{tabular}{l|c|c|c}
   Setting & 1 & 2 & 3 \\  \hline
   Run duration (s) & 1500 & 5000 & 5000 \\
   Number of runs & 111 & 111 & 110 \\
   Live time (days) & 1.93 & 6.42 & 6.37 \\
   Magnetic field & asymmetric & symmetric & symmetric \\
   $U_0$ (V) & \num{-18600} & \num{-18600} & \num{-18500} \\
   $\Updelta U_{\textrm{IE}}$ (V) & $\num{-5}$ & $\num{-5}$ & $\num{-100}$ \\
      \hline
     \end{tabular}
     \caption[Run settings]{Run settings used for the measurements described in this paper.  
     Settings 2 and 3 have lower electron rates and therefore require additional measurement time to get meaningful statistics.}
\label{table:RunSettings}
\end{table*}

A run is defined as a fixed length of time during which FPD data were collected.
In order to see the variations in the muon flux in each setting, runs were performed in a cyclic manner, iterating through each of the three settings.
This sequence was repeated automatically over the course of about 16 days, resulting in 113 completed cycles.
A small number of the runs had to be excluded from further analysis due to hardware issues.
The total measurement time for each setting is listed in \autoref{table:RunSettings}.

For all three settings, an acceleration voltage $U_{\textrm{PAE}}=+\SI{4}{\kilo\volt}$ was applied to the PAE, and a bias voltage $U_{\textrm{bias}}=+\SI{0.12}{\kilo\volt}$ was applied to the FPD wafer.
To determine the electron rate for a particular run, an electron region of interest (ROI) was defined using the initial electron energy (assumed to be \SI{\sim0}{\electronvolt} for production at the MS surface), the sum of the applied electrostatic potentials ($-U_0-\Updelta U_{\textrm{IE}}+U_{\textrm{PAE}}+U_{\textrm{bias}}$), and the energy resolution of the FPD. 
For the muon studies, the ROI is $\num{19.7}\num{-24.7}$ keV, and electrons with energies outside of this window were excluded from the data analysis.

\section{Validation of muon-induced background mechanism}
\label{sec:bkgmechanism}

Using FPD events that were coincident with those from the muon detectors, the time distribution for secondary electrons emitted from the MS surface was determined (\autoref{subsec:coincidenceAnalysis}).
The measured data was then compared to a simulated distribution in order to verify the model of muon-induced events (\autoref{subsec:comparisonSimulation}).
Finally, in \autoref{subsec:MuonCoincidenceNominal} we discuss the results of the electron-muon coincidence analysis under nominal magnetic field conditions.

\subsection{Coincidence analysis}
\label{subsec:coincidenceAnalysis}

A straightforward method to study the muon-induced background is to perform a coincidence analysis on muon and electron events. 
If muons passing through the MS vessel are responsible for creating electrons that reach the FPD system, one expects an excess of electron events in the time window following a muon event. 
(This is only true for the asymmetric magnetic field configuration; for the symmetric configuration, electrons can be trapped in the MS for long durations.)
The timing difference between muon and electron events allows the determination of the electron flight time, which can be compared with simulation.

In terms of event selection for a coincidence study with the FPD, it is desirable that all selected muons travel through the walls of the MS in order to have a chance of producing detectable electrons. 
Out of the available muon modules, modules 6, 7 and 8 are best suited to fulfill this condition (see \autoref{fig:MuonModuleLocations}).
The position and orientation of these modules relative to the MS is such that a muon that creates a signal in all three modules is geometrically constrained to have passed through the MS. 
(The deflection from the Lorentz force is negligible.)
Thus, only three-module muon events are used in the coincidence study, where such an event has concurrent signals within a \num{200}-\si{\nano\second} window.
This time window was chosen to account for the \num{50}-\si{\nano\second} timing resolution of the muon modules.

\begin{figure}
    \centering
    \includegraphics[width=\columnwidth]{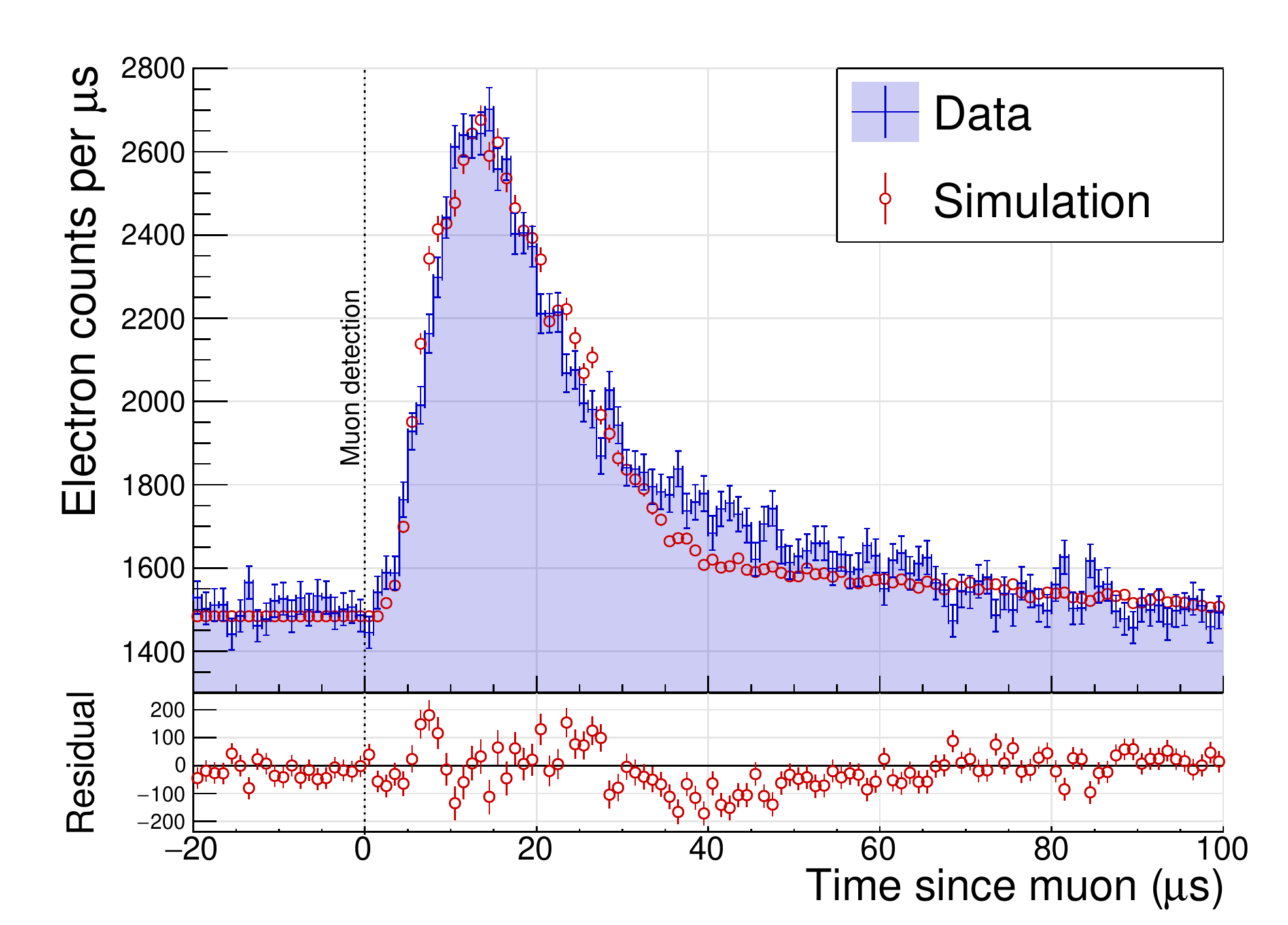}
    \caption[Muon-coincident electron events]{
    The distribution of the time differences between electron and muon events collected with field setting~1 (blue points and fill).  
    On the x-axis, $t=\SI{0}{\micro\second}$ corresponds to the detection of a three-module muon event.  
    Overlaid is the simulated time distribution (red markers) produced with \textsc{Kassiopeia} using the energy spectrum in \autoref{fig:EnergySpectrum} as input, with a maximum energy of \SI{50}{\eV}.  
     The simulated distribution is scaled to minimize the $\chi^2$/ndf (248.0/99 for $t > \SI{0}{\micro\second}$) with the offset from zero being fixed by the average electron counts prior to muon detection ($\SI{-100}{\micro\second} < t < \SI{0}{\micro\second}$).
    The error bars are purely statistical.  
    At the bottom of the figure, the residual $(\textrm{Simulation} - \textrm{Data})$ is displayed.}
        \label{fig:MuonSimulation}
     \end{figure}
     
In order to study events originating from the walls, only events from the outer 132 pixels of the detector were included in the analysis, since these pixels directly measure events from a well-defined section of the MS surface. 
For each electron event, the time difference between the electron event and the most recent muon event was tabulated, and the distribution of these time differences is shown in \autoref{fig:MuonSimulation} for the case of setting~1.
An excess above the random-coincidence level is clearly visible, indicating the presence of muon-coincident electron events.
The distribution peaks at time differences of about \SI{15}{\micro\second}.

\subsection{Comparison with simulation}
\label{subsec:comparisonSimulation}

To confirm that the time structure of the coincidence peak is consistent with the production of muon-induced electrons, Monte Carlo simulations were performed using \textsc{Kassiopeia}, the particle-tracking simulation package developed for the KATRIN experiment \cite{Furse2017}.
The simulation geometry included a simplified version of the system apparatus, consisting of the MS vessel and the FPD system, and employed the same electromagnetic field configuration used in setting~1, excluding the IE system.
\num{2e5} electrons were produced at the MS walls, uniformly spread over axial positions $\SI{-3.14}{\meter} \leq z \leq \SI{-0.27}{\meter}$, which is the range corresponding to the magnetic field lines that connect to the FPD (see \autoref{fig:MagneticSettings}).

True secondary electrons are produced isotropically inside the steel; electrons emitted from the surface, therefore, follow a cosine angular distribution~\cite{Henke1977,Seiler1983,Furman2002}.
This type of distribution was used to generate the starting angle of electrons in the simulation:
\begin{equation}
\frac{\text{d}n}{\text{d}\Omega}(\theta) \propto \cos\theta \Rightarrow
\frac{\text{d}n}{\text{d}\theta}(\theta) \propto \cos\theta\sin\theta.
 \label{eq:CosineLaw}
\end{equation}
Here, $\text{d}n$ is the number of emitted particles emitted in the solid angle $\text{d}\Omega$, and
$\theta$ is the angle between the particle momentum and the surface normal~\cite{Greenwood2002}.

\begin{figure}
  \centering
    \includegraphics[width=\columnwidth]{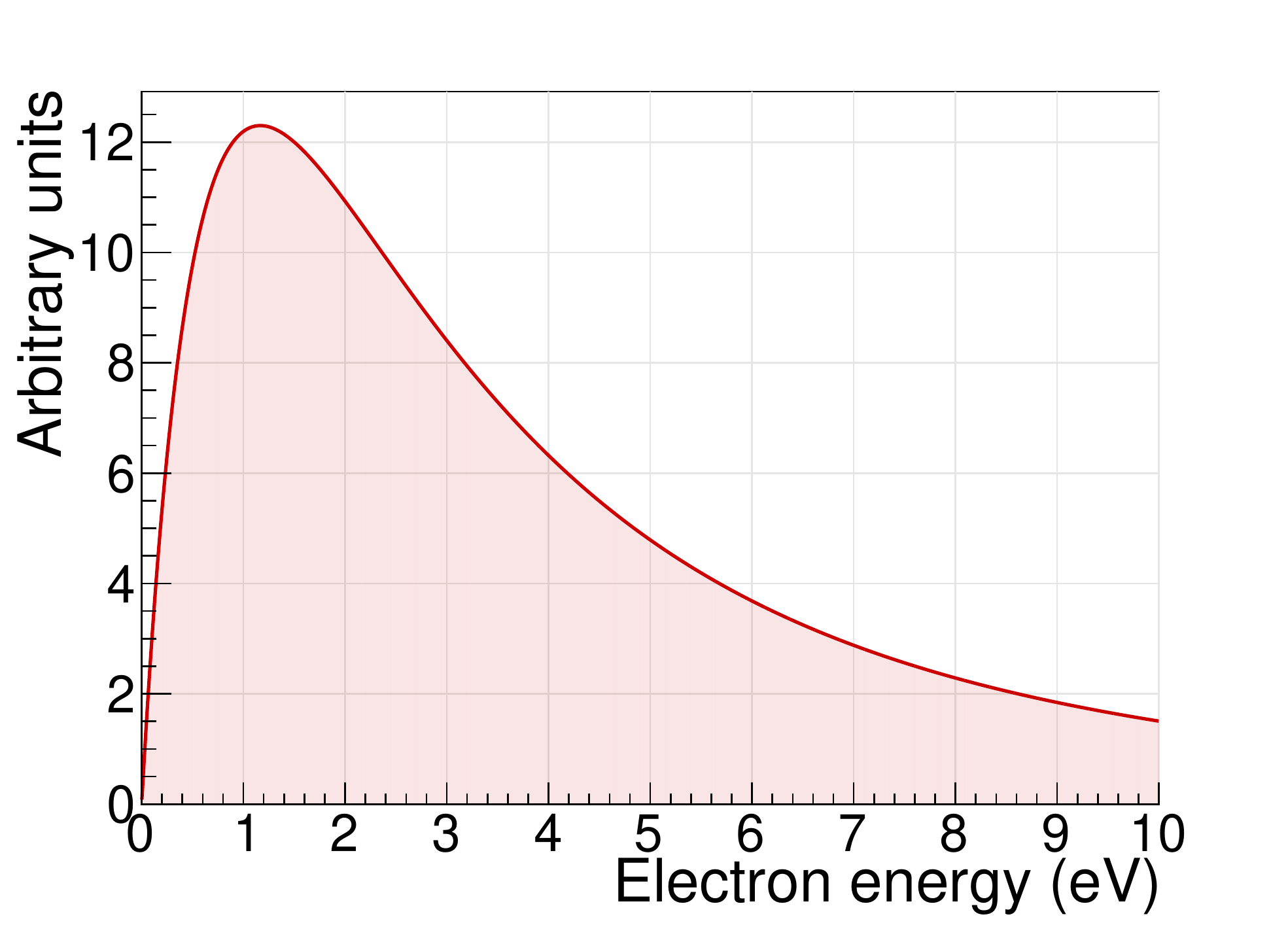}
    \caption[Secondary-electron spectra]{Secondary-electron energy spectrum, calculated from \autoref{eq:Spectrum} using a work function $\Phi = \SI{3.5}{\eV}$.}
     \label{fig:EnergySpectrum}
     \end{figure}

The energy spectrum $F$ of the simulated electrons was assumed to follow the theoretical shape for true secondaries emitted from a metal surface~\cite{Chung1974, Seiler1983, Joy2004}:
\begin{equation}
 F(E) = A \cdot \frac{E}{(E+\Phi)^{4}},
 \label{eq:Spectrum}
\end{equation}
where $E$ is the electron energy, $A$ is a normalization factor, and $\Phi$ is the work function.
The shape of the energy spectrum is independent of the incident muon since the muon's energy exceeds \SI{100}{eV}~\cite{Henke1977,Seiler1983}.
Transmission measurements with a photoelectron source~\cite{Behrens2017} previously found that the work function of the MS varied between \SI{3.39}{\eV} and \SI{3.65}{\eV} in the general timeframe of the muon studies~\cite{PhDBehrens2016}.
An energy spectrum with $\Phi = \SI{3.5}{\eV}$ was therefore utilized, which is plotted in \autoref{fig:EnergySpectrum}.

The flight times for the simulated electrons that reached the FPD are shown in \autoref{fig:MuonSimulation}.
The simulation replicates the essential features of the measured distribution of electron events.
However, at longer times ($t > \SI{15}{\micro\second}$) the simulation tends to underestimate the number of events. 
The simulation excludes any effects from IE system, which were placed at an offset voltage ($\Updelta U_{\textrm{IE}}$ = \SI{-5}{\V}) during the measurement.
This voltage is large enough to block a significant fraction of events from vessel walls.
However, secondary electrons are also emitted from the IE system and its holding structure (in the same way as from the walls), and these secondaries are not electrically shielded.
The combined effect of the blocked electrons from the walls and the additional events from the wire electrodes may explain the slight differences between the distributions.
Overall, however, the good agreement between measurement and simulation validates the proposed secondary-electron energy spectrum and confirms KATRIN's basic model of background production due to muons.

\subsection{Muon coincidence under nominal conditions}
\label{subsec:MuonCoincidenceNominal}

\begin{figure*}
 \centering
   \includegraphics[width=0.49\textwidth]{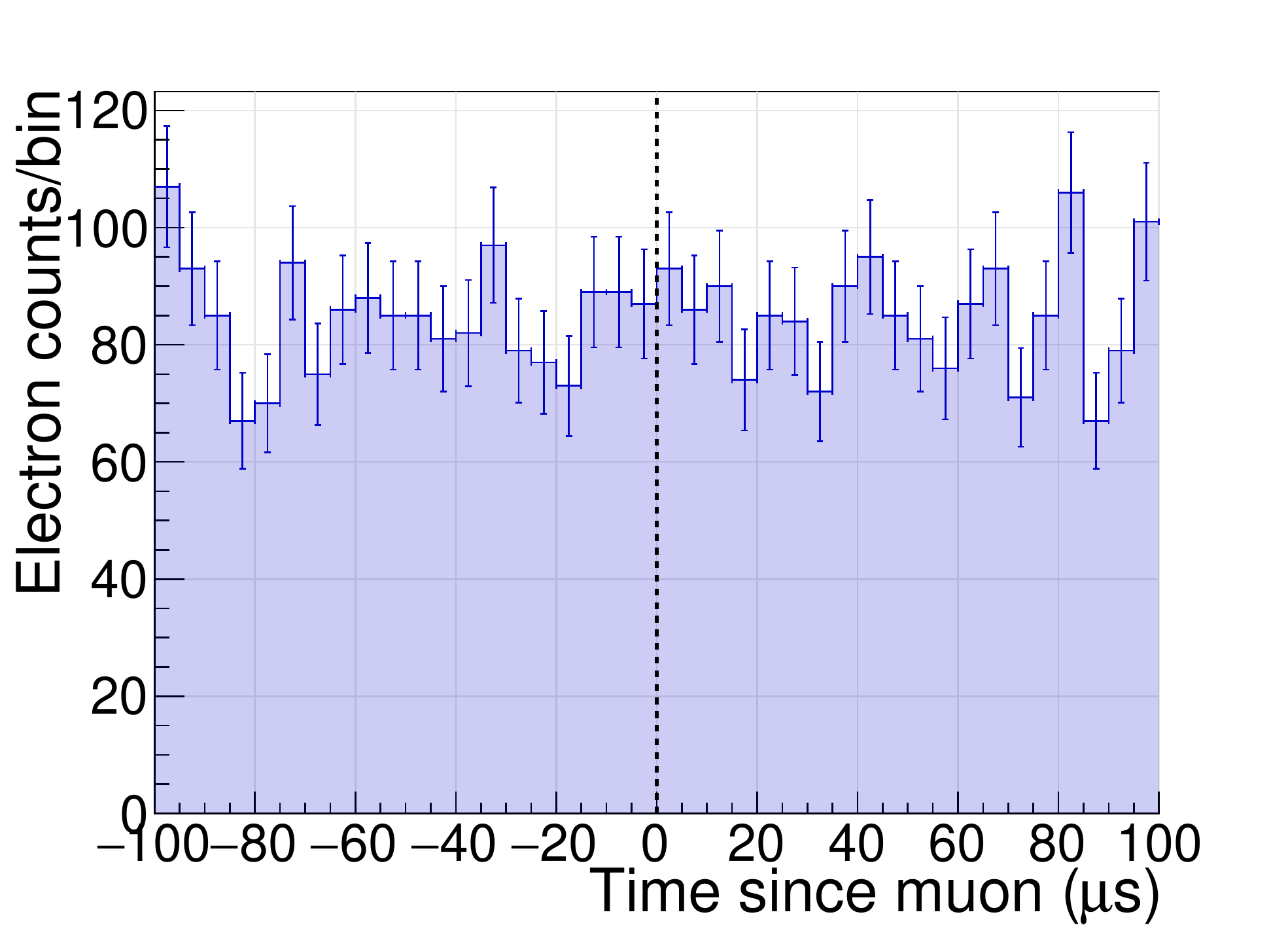}
  \includegraphics[width=0.49\textwidth]{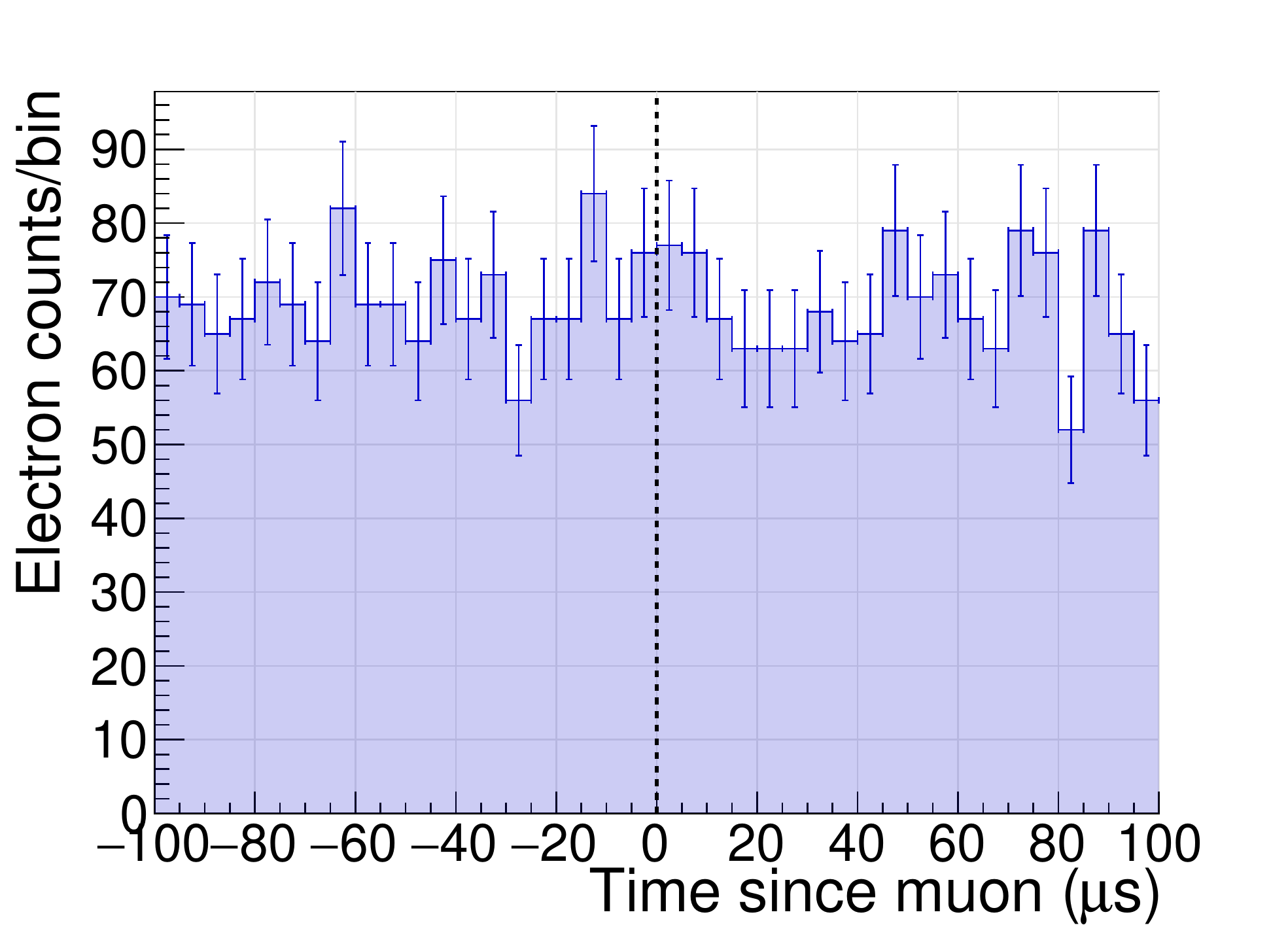}
 \caption{The distribution of the time differences between electron and muon events. 
 The black dashed line indicates $t=\SI{0}{\micro\second}$, the detection of a muon event.
 (\textit{Left}): Field setting~2.
 (\textit{Right}): Field setting~3.
 }
 \label{fig:CoincidenceSetting23}
\end{figure*}
     
The time distributions of muon-coincident electron events under setting~2 and setting~3 are displayed in \autoref{fig:CoincidenceSetting23}.  
No corresponding increase in the number of electron events following the three-module muon signals is observed.

One can attempt to set an upper limit on the muon-induced background rate by counting the excess number of events for $t>0$ compared with $t<0$, and then scaling the result appropriately to consider all muon events that pass through the MS, not just those that pass through modules 6, 7, and 8.
However, this approach is vulnerable to systematic uncertainties.
First, it is challenging to accurately extrapolate the coincidence rate for a particular region of the MS surface to the entire vessel without understanding the efficiency of electron transport as a function of the initial location on the MS surface.  
This requires significant particle-tracking simulations beyond the scope of the present paper.
A second difficulty is the possible time-dependent behavior of the secondaries.  
Electrons can be magnetically trapped in the symmetric magnetic field of setting 2 and setting 3 for up to several minutes~\cite{PhDWandkowsky2013}; thus, muon-induced secondaries and any additional electrons produced in the trap can reach the FPD well beyond the \SI{100}{\micro\second} interval applied in the coincidence study.

To test the statistical sensitivity of using the coincidence data to set an upper limit, a naive extrapolation to the entire MS was performed; the resulting upper limit on the muon-induced rate is comparable to the value derived from the correlation study (see \autoref{sec:residualbkg}).
Because the uncertainties for the coincidence approach are difficult to calculate, this method was not developed further.

\section{Correlation of cosmic-ray muon rate with detected background}
\label{sec:correlationanalysis}

For the correlation analysis of the background electron rate and the muon rate, the following assumptions were made: the background consists of a fluctuating muon-induced component and a constant component of at least one other source; and the muon-induced background component is directly correlated with the muon rate.
These assumptions lead to the following formula for the background electron rate $R_{\textrm{e}}(t)$:
\begin{equation}
 R_{\textrm{e}}(t) = K \cdot R_{\upmu}(t) + R_{\textrm{x}},
 \label{eq:bgrate}
\end{equation}
where $R_{\upmu}(t)$ is the muon rate measured by the muon modules, $R_{\textrm{x}}$ is the constant background component, and $K$ is the coefficient representing the linear relation between the muon rate and the resulting rate of secondary electrons detected by the FPD. 

Translating this into normalized rates, \autoref{eq:bgrate} becomes:
\begin{equation}
  \frac{R_{\textrm{e}} (t)}{\overline{R_{\textrm{e}}}} = \underbrace{K \cdot \frac{\overline{R_{\upmu}}}{\overline{R_{\textrm{e}}}}}_{=:m}
  \cdot \frac{R_{\upmu} (t)}{\overline{R_{\upmu}}} + \underbrace{\frac{R_{\textrm{x}}}{\overline{R_{\textrm{e}}}}}_{=1-m}
  = m \cdot \frac{R_{\upmu} (t)}{\overline{R_{\upmu}}} + (1-m).
   \label{eq:correlationEquation}
\end{equation}
with $\overline{R_{\textrm{e}}}$ and $\overline{R_\upmu}$ being the mean electron and muon rate, respectively.
The only unknown parameter is $m$, which represents the fraction of background that is muon-induced.
Plotting the normalized electron rate as a function of the normalized muon rate, $m$ is given by the slope.

The correlation analysis was performed for all three field settings; a summary of the results is given in \autoref{table:correlationSummary}.
The correlation under asymmetric field setting is described in \autoref{subsec:CorrelationAsymmetric}, and in \autoref{subsec:AlphaCalculation} the measured muon-induced fraction is used to determine the production rate of muon-induced secondary electrons in the MS.
The analysis of the symmetric field correlation data is described in \autoref{sec:residualbkg}.

\begin{table*}
  \centering
  \begin{tabular}{cc|c|c|cc}
   \hline
   Setting & Selection & FPD Rate (\si{\cps}) & Muon Rate (\si{\cps}) & Correlation $r$ & Slope $m$ \\
   \hline
   \multirow{4}{*}{1} & all events & \hspace{0pt} \num{252.726 \pm 0.068} & \multirow{4}{*}{\num{1413.14 \pm 0.09}} & \hspace{5pt} \num{0.70 \pm 0.06} & \hspace{5pt} \num{0.123 \pm 0.012} \\
   & $M=1$ & \hspace{0pt} \num{112.787 \pm 0.026} & & \hspace{5pt} \num{0.90 \pm 0.03} & \hspace{5pt} \num{0.225 \pm 0.010} \\
   & $M=2$ & \hspace{5pt} \num{55.129 \pm 0.026} & & \hspace{5pt} \num{0.41 \pm 0.08} & \hspace{5pt} \num{0.098 \pm 0.021} \\
   & $M\geq3$ & \hspace{5pt} \num{84.817 \pm 0.054} & & \hspace{5pt} \num{0.02 \pm 0.08} & \hspace{5pt} \num{0.005 \pm 0.029} \\
   \hline
   2 & all events & \hspace{10pt} \num{0.8259 \pm 0.0015} & \num{1421.15 \pm 0.05} & \num{-0.02 \pm 0.10} & \num{-0.013 \pm 0.079} \\
   3 & all events & \hspace{10pt} \num{0.6639 \pm 0.0014} & \num{1420.69 \pm 0.05} & \hspace{5pt} \num{0.12 \pm 0.08} & \hspace{5pt} \num{0.118 \pm 0.093} \\
  \hline
  \end{tabular}
   \caption[Summary of correlation results]{Summary of the electron-muon rate correlation results. 
   The rates are listed in units of counts per second (cps).
      $M$ is the multiplicity of the electron event.
   The FPD and muon rates are the average values over the measurement campaign.}
    \label{table:correlationSummary}
\end{table*}


\subsection{Correlation under asymmetric magnetic field conditions}
\label{subsec:CorrelationAsymmetric}

\begin{figure}
 \centering
 \includegraphics[width=\columnwidth]{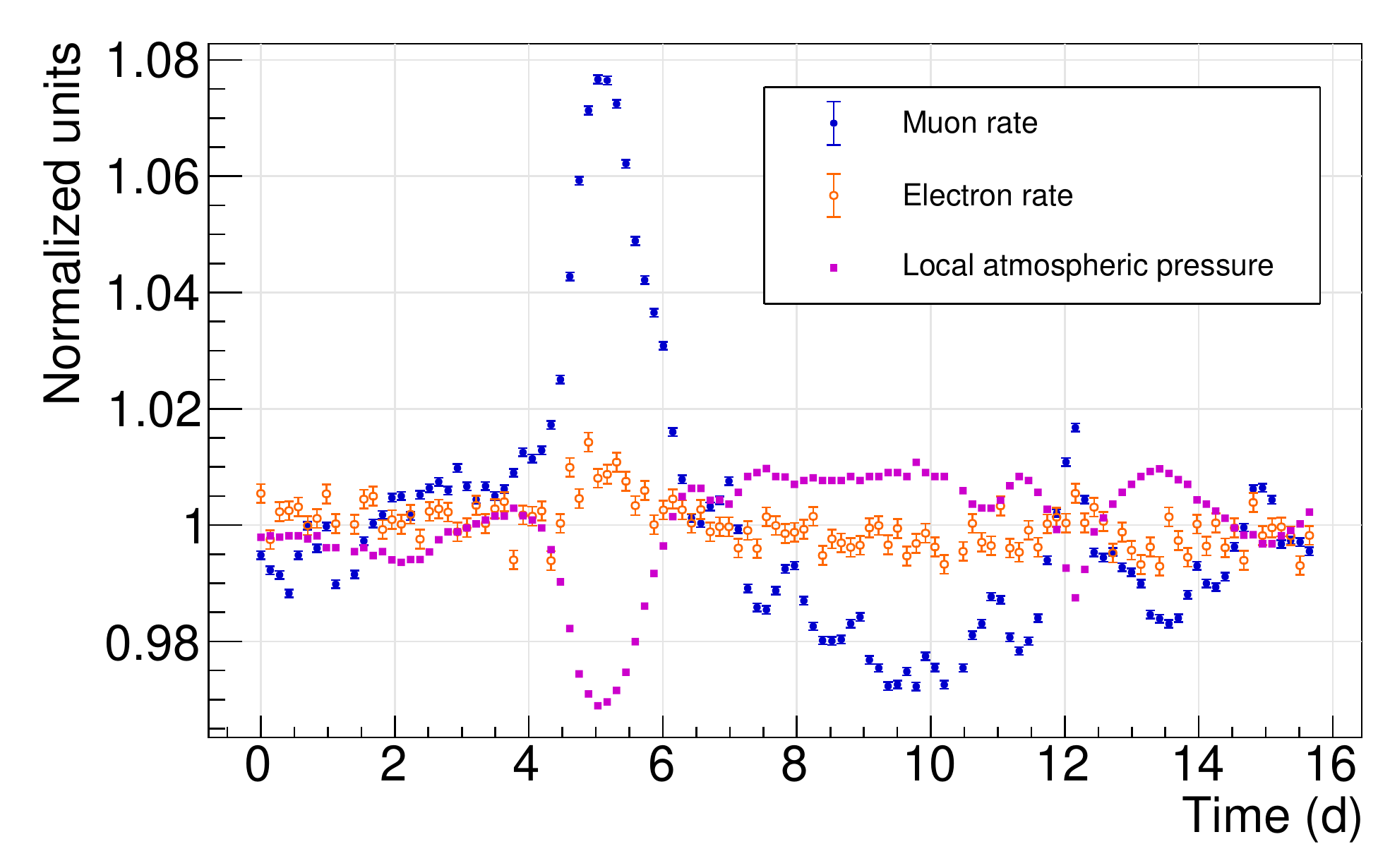}
 \caption{The normalized muon and electron rates as a function of time under setting 1.
 Each datapoint corresponds to the average value during an FPD run.
 The pressure was measured by a weather station housed in the spectrometer hall (located at ground level), which correlates with the atmospheric pressure.}
 \label{fig:normRates}
\end{figure}

The normalized muon and electron rates as functions of time for setting 1 are displayed in \autoref{fig:normRates}.
A large increase in the muon rate is visible near day 5, caused by a low-pressure weather system that passed over the experiment.
The reduced atmospheric pressure indicates a decreased air density (and therefore a larger mean free path) that results in more muons reaching the Earth's surface~\cite{Abrahao2016}.

\begin{figure}
 \centering
 \includegraphics[width=\columnwidth]{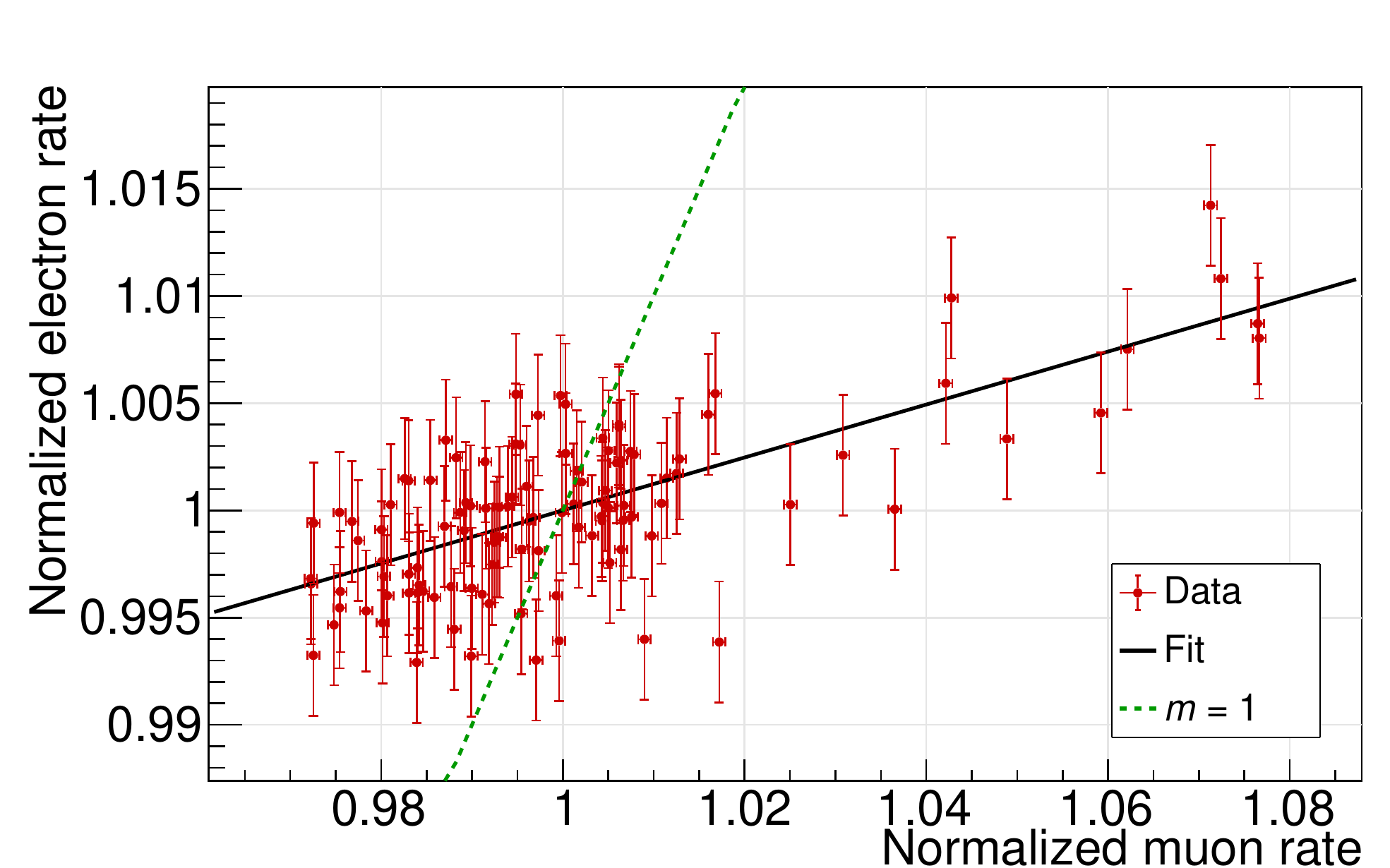}
 \caption{Correlation of the electron rate and the muon rate with an asymmetric magnetic field (setting~1), where electrons from the MS surface are guided to the FPD. 
Each data point represents a single FPD run.
 The correlation coefficient is $r=0.70 \pm 0.06$, whereas the slope (solid black line) shows that only a fraction $m = \num{0.123 \pm 0.012}$ of the background is muon-induced. The case of a completely muon-induced background is shown with the green dashed line for comparison.}
 \label{fig:setting1}
\end{figure}

In \autoref{fig:setting1}, the normalized muon and electron rates are plotted against each other, and the fit to \autoref{eq:correlationEquation} is also shown. 
The fraction of muon-induced background is \num{0.123 \pm 0.012}. 
This result indicates that muons make up a sizable fraction but not the majority of secondary electron events originating from the MS surface and IE system.
(Because of the electrostatic shielding potential applied during setting~1, a significant portion of the background from low-energy electrons originates from the IE system.)
The Pearson correlation coefficient $r$ was calculated to be \num{0.70 \pm 0.06}, which indicates significant linear correlation. 
The uncertainty was estimated via a case resampling Monte Carlo bootstrap method~\cite{efron1994}.

\begin{figure*}

\centering
   \includegraphics[width=0.49\textwidth]{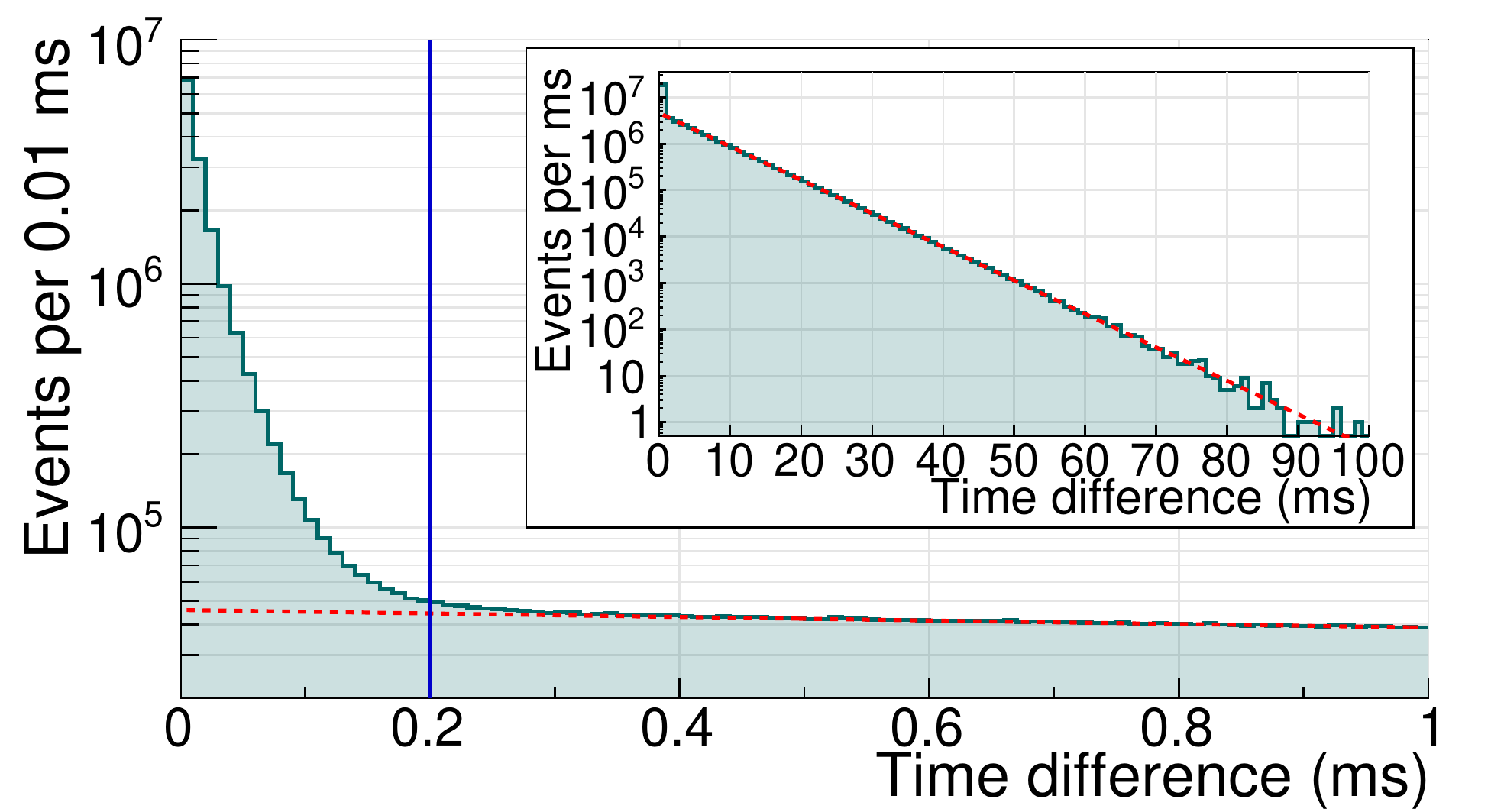}
  \includegraphics[width=0.49\textwidth]{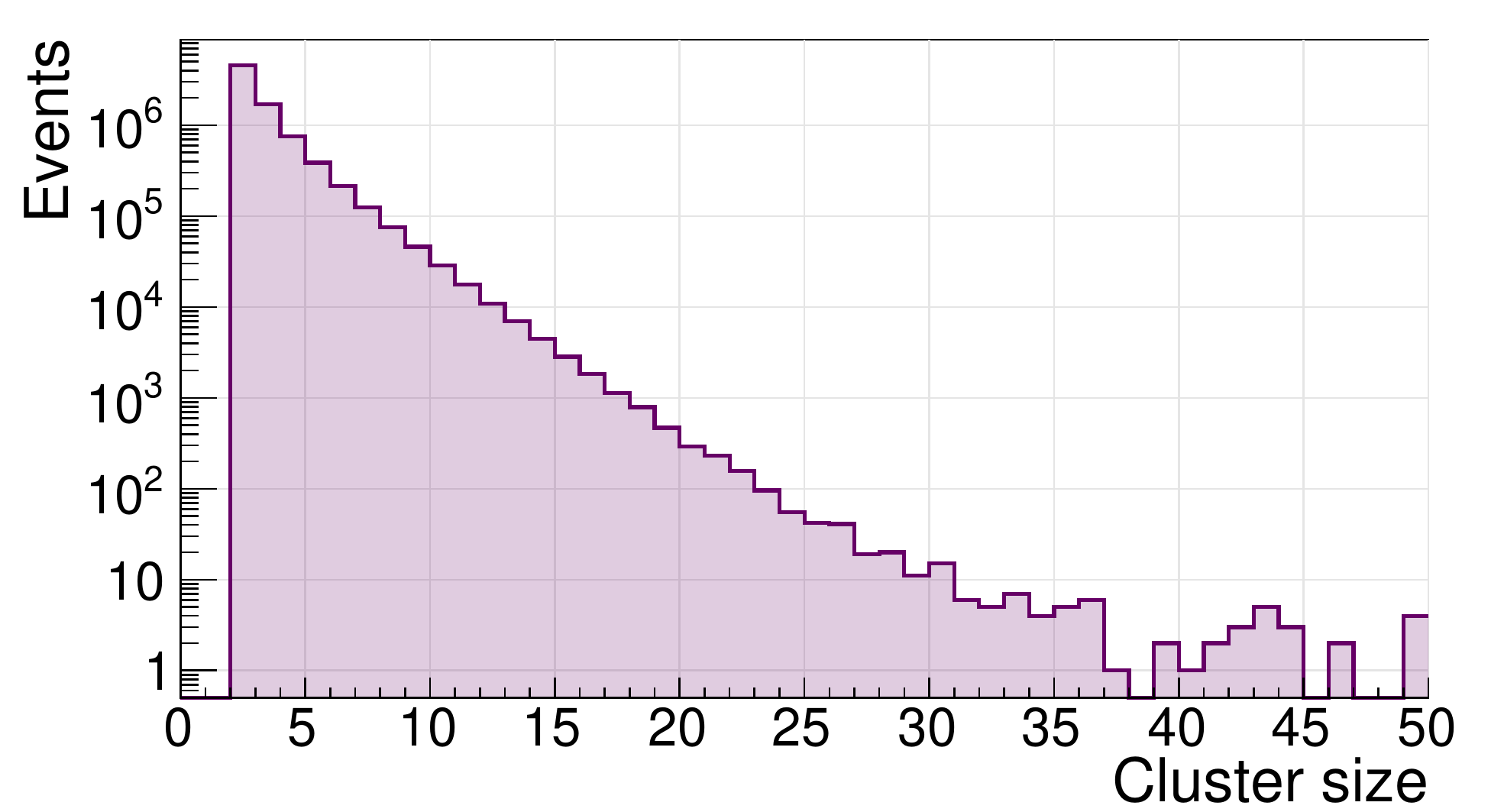}
 \caption{
 (\textit{Left}): Distribution of the time difference between electron events for measurement setting~1.  
    Below \SI{0.2}{\ms} (shown by the solid blue line), there is a change in slope, indicating a contribution from correlated (i.e. clustered) events.
 The inset shows the time distribution for differences up to \SI{100}{\ms}.
  A fit to the distribution for $\SI{1}{\ms} < \Updelta t < \SI{100}{\ms}$ is given by the red dashed line. 
 (\textit{Right}): Multiplicity distribution for the cluster events ($\Updelta t < \SI{0.2}{\milli\second}$), excluding cluster sizes greater than 50.
 }
 \label{fig:interarrival}
 \end{figure*}

The distribution of the time difference $\Updelta t$ between electron events for setting~1 is shown in \autoref{fig:interarrival}, left panel.
It can be seen that there are a large number of events with time difference $\Updelta t$ less than \SI{0.2}{\milli\second}.
At longer time differences, however, the distribution has a constant slope.
The distribution in the figure can be explained by two processes with different multiplicity distributions for secondary-electron production. 
One process with multiplicity $M=1$ produces a single electron that is detected at the FPD. 
These events are called ``single'' events, and they are Poisson-distributed, contributing to the constant slope at large time differences.

Another process creates clustered electron events ($M\geq2$) which arrive at the FPD within short time intervals (\autoref{fig:interarrival}, right panel). 
The electrons produced from these high-multiplicity events will have a spread of initial energies and pitch angles, resulting in the observed flight time differences of up to \SI{0.2}{\milli\second}. 
These events are referred to as ``cluster'' events.
The multiplicity $M$ of a cluster is defined using a sliding time window of duration $d = \SI{0.2}{\milli\second}$.  
All events that fall within $d$ of neighboring events are grouped together, such that the time difference between the first and last event in the group can in principle be greater than $d$. 
The event multiplicity is defined as the number of events in the group.
Using this criterion, about \SI{45}{\%} of the electron events in setting~1 are classified as single events, with the remainder being cluster events.
 Due to the presence of cluster events, the FPD event rate is non-Poissonian.
All FPD rates used in the correlation analyses (and shown in the figures) utilize the RMS error.
 For the muon rates, the given errors are purely statistical.
 
\begin{figure*}
  \centering
  \includegraphics[width=0.49\textwidth]{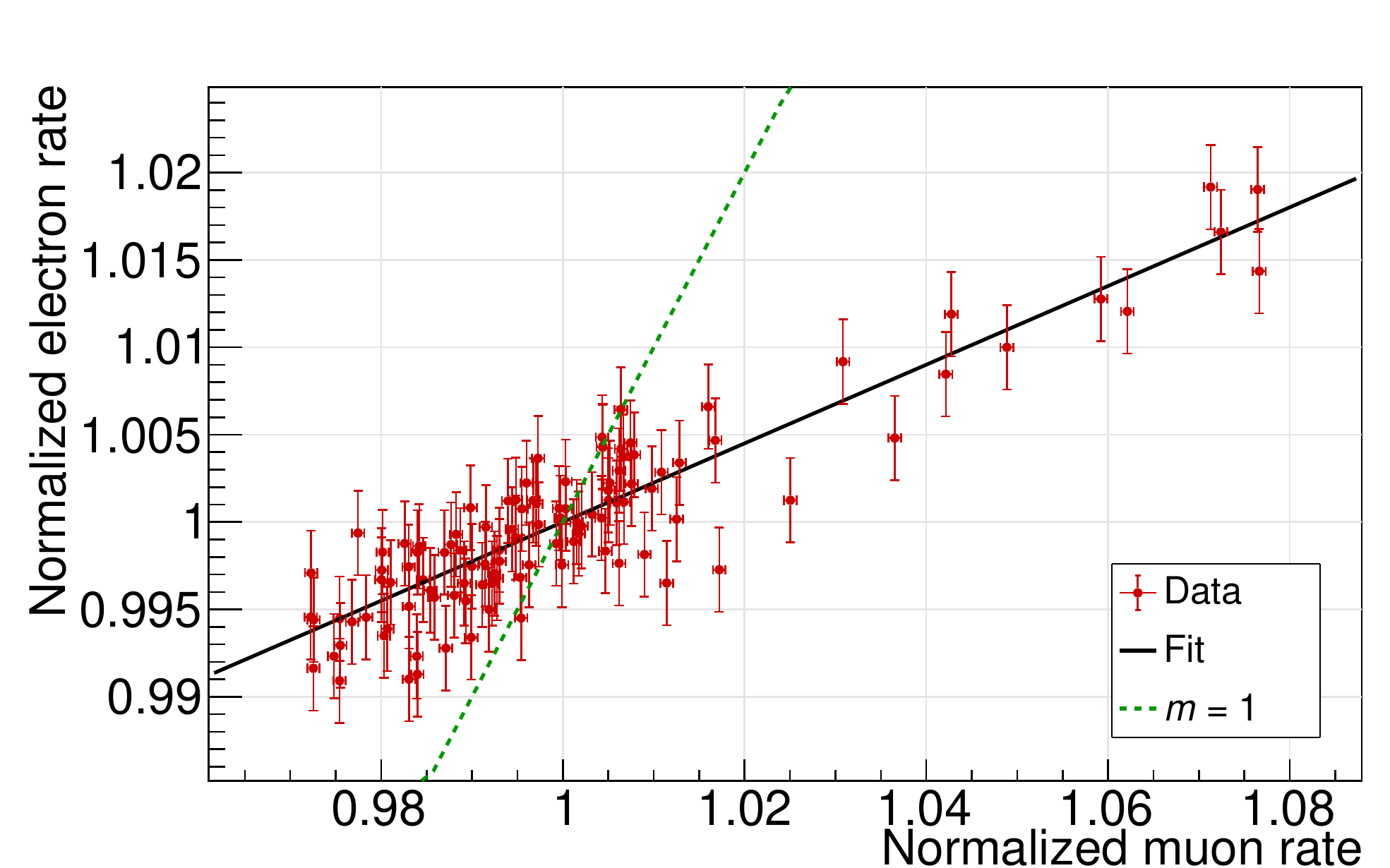}
  \includegraphics[width=0.49\textwidth]{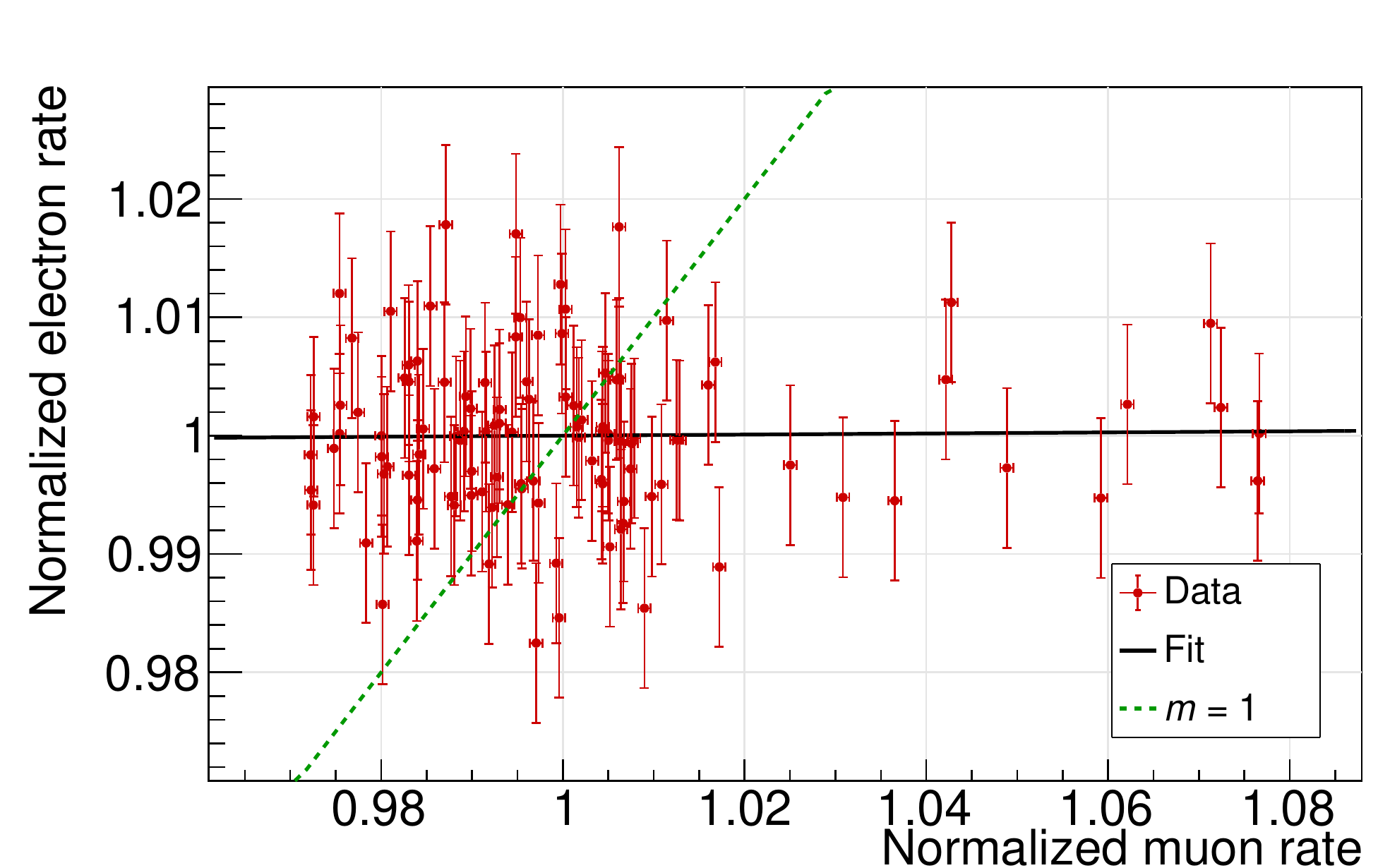}
  \caption[Correlation single and cluster events]{ 
  (\textit{Left}): Single electron rate ($M$=1) as a function of muon rate for setting~1.  
  There is strong correlation ($r = 0.90 \pm 0.03$), and the muon-induced fraction $m$ is \num{0.225 \pm 0.010}.
(\textit{Right}): Cluster electron rate ($M$$\geq$3) as a function of muon rate for setting~1, showing no correlation ($r = 0.02 \pm 0.08$).
For both plots, the green dashed line represents the case of fully muon-induced background.}
  \label{fig:ClusterSingle}
  \end{figure*}

The correlation analysis was repeated for different electron event multiplicities (see \autoref{table:correlationSummary}).
The single ($M=1$) electron event rate shows a strong correlation with the muon rate (\autoref{fig:ClusterSingle}, left panel).
A weaker correlation is observed for the double ($M=2$) electron event rate. 
It should be mentioned, however, that a portion of the cluster event rate comes from ``accidental'' clustering--single events that statistically happen to fall within $\Updelta t$ of another single event or another cluster.
Thus, the measured correlations and slopes for $M>2$ are not corrected for the contribution from single events.
Nevertheless, no significant correlation is found for cluster events with $M\geq3$ (\autoref{fig:ClusterSingle}, right panel).
This result strongly indicates that muons dominantly produce events with small multiplicities.  

  \begin{figure*}
    \centering
  \includegraphics[width=0.49\textwidth]{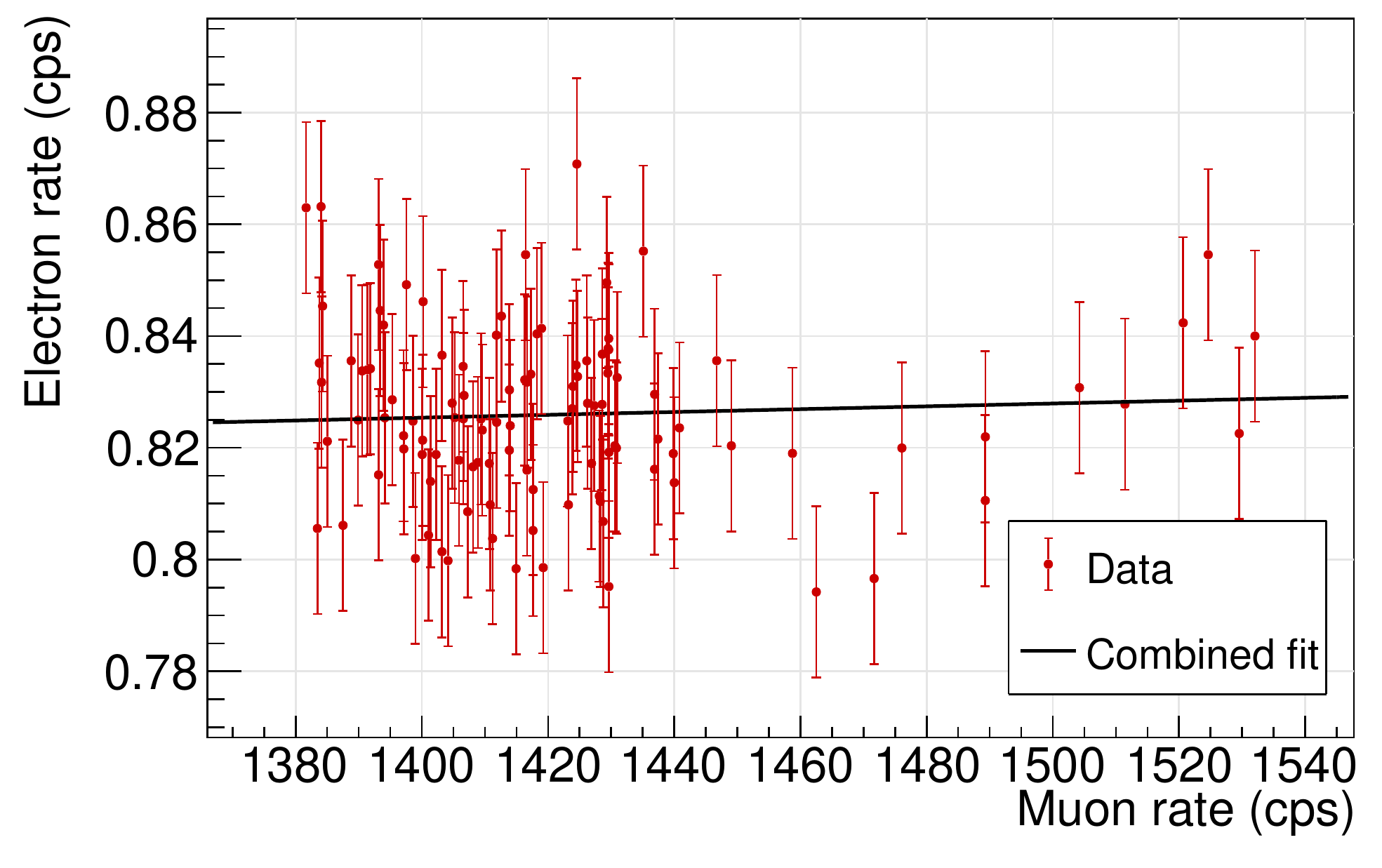}
  \includegraphics[width=0.49\textwidth]{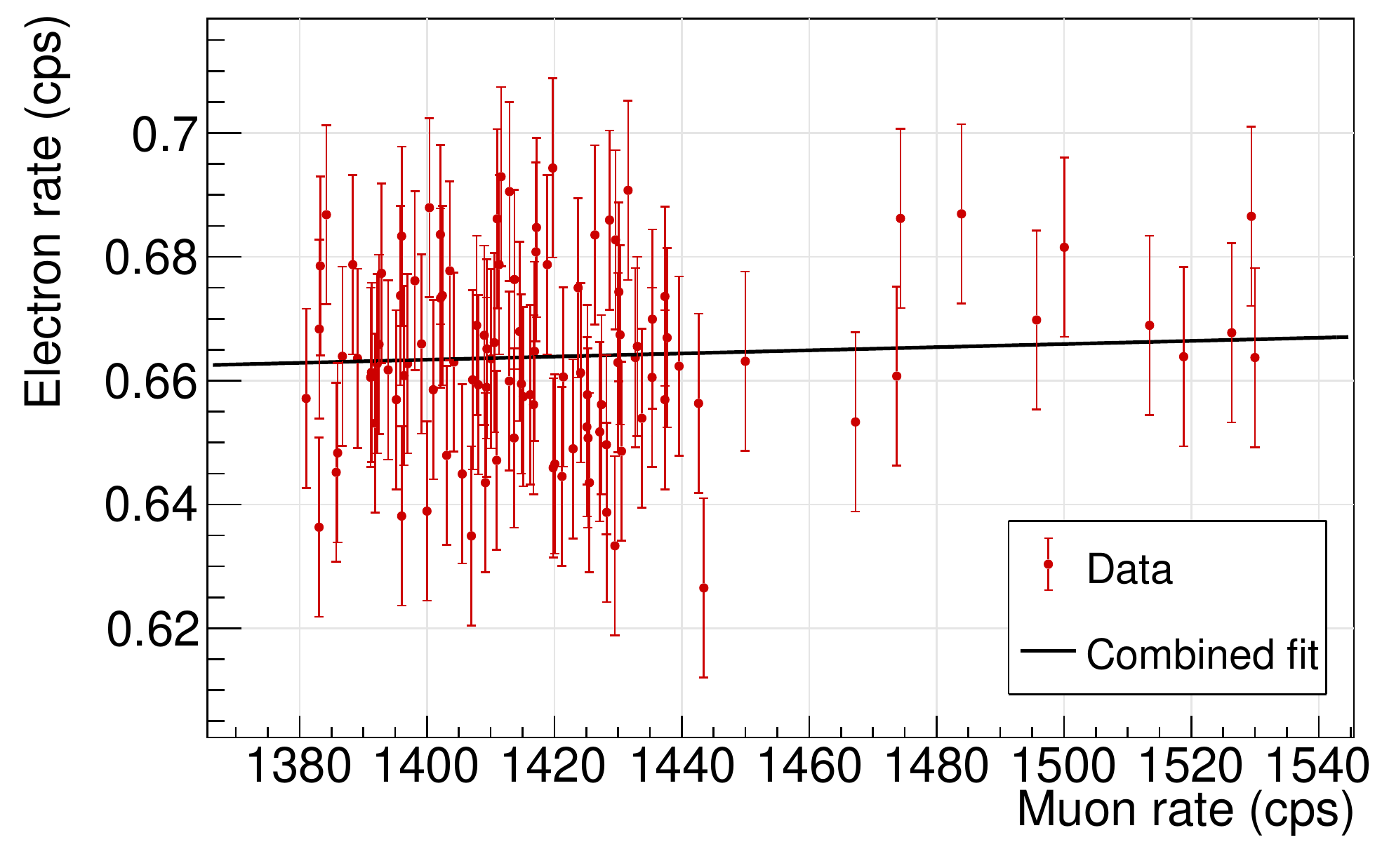}
  \caption{Electron rate as a function of muon rate for setting 2 (left) and setting 3 (right).
A simultaneous linear fit of both datasets (black line) finds a slope $K=\num{2.5 \pm 3.2 e-5 }$, indicating $m = \num{0.044 \pm 0.054}$ for setting 2 and $m = \num{0.054 \pm 0.068}$ for setting 3.}
  \label{fig:setting23}
  
\end{figure*}

\subsection{Electron production rate from muons}
\label{subsec:AlphaCalculation}

Knowing the value of $m$ for setting~1, it is possible to make a rough determination of the electron production rate by a single muon crossing the MS surface.
This quantity, which we denote as $\alpha$, can be obtained from the following equation:
\begin{equation}
 \alpha = \frac{m \cdot N_\text{FPD} \cdot C}{N_\upmu}.
 \label{eq:alpha}
\end{equation}
The numerator gives the number of muon-induced electrons emitted from the inner surface.
$N_\text{FPD}$ is the rate of electrons from the MS surface that reach the FPD for the same magnetic field configuration as setting~1, but without electrostatic shielding (i.e. with $\Updelta U_{\textrm{IE}}=0$).  
A measurement, described elsewhere~\cite{PhDHarms2015}, found this value to be \num{790} counts per second (cps).
$C$ is a correction factor that accounts for the probability of an electron emitted from the surface to be detected by the FPD.
From \textsc{Kassiopeia} simulations (see \autoref{subsec:comparisonSimulation}), it was determined that electrons have an \SI{13.2}{\%} chance of reaching the FPD when averaging over the secondary-electron energy spectrum.
Considering that the FPD itself has a detection efficiency of about \SI{95}{\%}~\cite{Amsbaugh2015}, then $C \approx (0.95\cdot0.132)^{-1}=8.0$.

The denominator, $N_\upmu$, is the rate of muon hits on the MS surface.
From the \textsc{Geant4} simulation mentioned in \autoref{sec:introduction}, $N_\upmu$ was determined to be \SI{13.3}{\kilo\cps} for the portion of the MS surface measured with setting~1.
Applying the aforementioned values to \autoref{eq:alpha}, one finds that $\alpha\approx\num{0.058}$.
This result indicates that one secondary electron is emitted from the MS surface for about every 17 muon crossings.

As mentioned previously, the total muon flux through the MS is about \num{4.5e4}~$\upmu$/\si{\second}.
Multiplying this number by $\alpha$ (and a factor of 2 to account for each muon making two crossings of the inner MS surface), one finds a rate of \num{5.3e3} muon-induced secondary electrons per second.
With muons responsible for \SI{12}{\%} of the total rate from the surface, about \num{4.3e4} electrons in total are emitted from the MS surface every second.
This result highlights the importance of using magnetic and electrostatic shielding to prevent these electrons from reaching the FPD.

\section{Residual muon-induced background with symmetric magnetic field}
\label{sec:residualbkg}

Turning to settings 2 and 3, no significant correlation was found between the muon and FPD rates (see \autoref{table:correlationSummary}).
In addition, performing a single/cluster event analysis is not useful in this case since the FPD events for these settings are essentially all single events ($>\SI{99}{\%}$).

The measured muon-induced fractions for setting~2 and setting~3 are consistent with zero.
For setting~3, which is closest to the nominal KATRIN operating mode, assuming Gaussian errors and constraining $m$ to be non-negative~\cite{Feldman1997} leads to an upper limit on the true muon-induced fraction of $m < 0.27~(\SI{90}{\%}~\mathrm{C.L.})$.
However, it is possible to reduce this upper limit by doing a simultaneous fit of the setting~2 and setting~3 data.
Setting~2 has reduced shielding and should therefore have a larger muon-induced background component, if such a background is indeed present.
By performing a simultaneous fit of the two datasets, one naively expects to raise the measured upper limit on the muon-induced background compared with an analysis with only setting~3, but the opposite effect is observed since the analysis is statistics-limited.
(Although the correlation $r$ should be larger for setting~2 compared with setting~3, this cannot be seen due to the large uncertainties on the correlation coefficients.)

\autoref{fig:setting23} shows the simultaneous fit for setting~2 and setting~3, which results in a value of $m=\num{0.054 \pm 0.068}$ for setting~3.
Following the unified approach~\cite{Feldman1997}, the upper limit on the muon-induced fraction is
\begin{equation}
m < 0.166 \quad (\SI{90}{\%}~\mathrm{C.L.}).
 \label{eq:UpperLimit2}
\end{equation}
The average background rate under setting 3 is \SI{0.692}{\cps}, after applying a corrective factor (148/142) to account for the six excluded detector pixels.
Thus, \autoref{eq:UpperLimit2} indicates that cosmic-ray muons contribute less than \SI{0.115}{\cps} to the total background rate.


A separate study was performed to check the sensitivity of the correlation analysis, using an ensemble of toy measurements generated based on the observed muon and electron rates.
For a measurement with the duration reported in this paper, we calculated a \SI{95}{\%} probability of detecting $m=0.10$ for a simultaneous fit of data from setting 2 and setting 3, which indicates that the correlation analysis is sensitive to $m>0.10$.
Thus, the measured null result for the muon-induced fraction is consistent with the expected statistical sensitivity of the measurement.

\section{Conclusion}
\label{sec:conclusion}

In order to reach KATRIN's design sensitivity, it is necessary to have a good understanding of the background processes inside the MS, including muon-induced backgrounds.
Using an electromagnetic configuration in which electrons are directly guided from the surface of the MS to the FPD, rate correlations with the muon detector system indicate that \SI{12.3 \pm 1.2}{\%} of the observed rate from the MS surface is muon-induced.
In addition, the fraction of single events that are muon-induced appears to be significantly higher  than the fraction of muon-induced cluster events.
Although not discussed in this paper, the remaining \SI{88}{\%} of electrons emitted from the surface are created from several sources, including environmental gamma radiation~\cite{GammaPaper} and ionization processes caused by the decay of $^{210}$Pb on the inner surface of the MS vessel~\cite{PhDHarms2015}.

A rough calculation indicates that, on average, one secondary electron is produced for every 17 muons passing through the MS.
However, the magnetic shielding of the KATRIN flux tube is highly effective at mitigating this background. 
In an electromagnetic configuration similar to that planned for neutrino-mass measurements, there is no correlation between the muon rate and the rate of detected electrons. 
An analysis of all data with magnetic shielding indicates that cosmic-ray muons are responsible for less than \SI{16.6}{\%} of the overall MS background rate, at \SI{90}{\%} confidence.
This corresponds to an upper limit of \SI{0.115}{\cps} for a total background rate of \SI{0.692}{\cps}.
Muon-induced backgrounds are therefore not a significant concern for KATRIN, although they may become more important as other background sources are alleviated. 

A significant potential source of background electrons is due to the decay of radon in the MS vessel; however, the installation of liquid-nitrogen cooled baffles between the MS volume and the NEG pumps has effectively mitigated this background process~\cite{Drexlin2017, RadonPaper}.
In the current background model for KATRIN, the largest background contribution originates from the ionization of Rydberg atoms, produced from the decay of $^{210}$Pb on the surface of the MS vessel~\cite{PhDHarms2015}.  
Additional details regarding this background source can be found in~\cite{Fraenkle2017}.

The KATRIN signal rate from $\upbeta$-decay electrons is highly dependent on the MS retarding potential $U_0$, but it is expected to be on the order of \SI{0.1}{\cps} within the last \SI{10}{\eV} of the $\upbeta$-spectrum~\cite{arXivKleesiek2018}.
The design sensitivity to the neutrino mass assumes a background rate of \SI{0.01}{\cps}~\cite{KATRIN2005}.
Although the present background level is roughly 50 times larger than this goal~\cite{Fraenkle2017}, the sensitivity does not scale simply with the background level.
By adjusting the measuring time distribution (i.e. the range of $U_0$ and the time spent at each value)~\cite{arXivKleesiek2018} as well as optimizing the magnetic field configuration to reduce the radial-dependent background from the MS volume~\cite{PhDHarms2015}, it is possible to mitigate the effect of the large background rate and reach a sensitivity to the neutrino mass of \SI{0.24}{\electronvolt\per c\squared} (\SI{90}{\percent} C.L.)~\cite{Fraenkle2017}.
Nonetheless, in order to improve the sensitivity, further investigations of background processes in the MS are ongoing.

\section*{Acknowledgements}
\label{sec:Acknowledgements}
We acknowledge the support of Helmholtz Association (HGF), Ministry for Education and Research BMBF (5A17PDA, 05A17PM3, 05A17PX3, 05A17VK2, and 05A17WO3), Helmholtz Alliance for Astroparticle Physics (HAP), and Helmholtz Young Investigator Group (VH-NG-1055) in Germany; Ministry of Education, Youth and Sport (\text{CANAM-LM2011019}), cooperation with the JINR Dubna (3+3 grants) 2017--2019 in the Czech Republic; and the Department of Energy through grants DE-FG02-97ER41020, DE-FG02-94ER40818, \text{DE-SC0004036}, DE-FG02-97ER41033, DE-FG02-97ER41041, DE-AC02-05CH11231, \text{DE-SC0011091}, and \text{DE-SC0019304} in the United States.



\bibliographystyle{elsarticle-num} 
\bibliography{BibMuonPaper}

\begin{thebibliography}{10}
\expandafter\ifx\csname url\endcsname\relax
  \def\url#1{\texttt{#1}}\fi
\expandafter\ifx\csname urlprefix\endcsname\relax\def\urlprefix{URL }\fi
\expandafter\ifx\csname href\endcsname\relax
  \def\href#1#2{#2} \def\path#1{#1}\fi

\bibitem{Wendell2010}
R.~Wendell, et~al., Atmospheric neutrino oscillation analysis with subleading
  effects in {Super-Kamiokande} {I}, {II}, and {III}, Phys. Rev. D 81 (2010)
  092004.
\newblock \href {http://dx.doi.org/10.1103/PhysRevD.81.092004}
  {\path{doi:10.1103/PhysRevD.81.092004}}.

\bibitem{Aharmim2013}
B.~Aharmim, et~al., Combined analysis of all three phases of solar neutrino
  data from the {Sudbury Neutrino Observatory}, Phys. Rev. C 88 (2013) 025501.
\newblock \href {http://dx.doi.org/10.1103/PhysRevC.88.025501}
  {\path{doi:10.1103/PhysRevC.88.025501}}.

\bibitem{Kraus2005}
C.~Kraus, et~al., Final results from phase {II} of the {Mainz} neutrino mass
  search in tritium {$\beta$} decay, Eur. Phys. J. C 40~(4) (2005) 447--468.
\newblock \href {http://dx.doi.org/10.1140/epjc/s2005-02139-7}
  {\path{doi:10.1140/epjc/s2005-02139-7}}.

\bibitem{Lobashev2003}
V.~Lobashev, {The search for the neutrino mass by direct method in the tritium
  beta-decay and perspectives of study it in the project KATRIN}, Nucl. Phys. A
  719 (2003) 153--160.
\newblock \href {http://dx.doi.org/10.1016/S0375-9474(03)00985-0}
  {\path{doi:10.1016/S0375-9474(03)00985-0}}.

\bibitem{Beamson1980}
G.~Beamson, et~al., The collimating and magnifying properties of a
  superconducting field photoelectron spectrometer, J. Phys. E: Sci. Inst.
  13~(1) (1980) 64.
\newblock \href {http://dx.doi.org/10.1088/0022-3735/13/1/018}
  {\path{doi:10.1088/0022-3735/13/1/018}}.

\bibitem{Lobashev1985}
V.~Lobashev, P.~Spivak, A method for measuring the electron antineutrino rest
  mass, Nucl. Instr. and Meth. A 240~(2) (1985) 305--310.
\newblock \href {http://dx.doi.org/10.1016/0168-9002(85)90640-0}
  {\path{doi:10.1016/0168-9002(85)90640-0}}.

\bibitem{Picard1992}
A.~Picard, et~al., A solenoid retarding spectrometer with high resolution and
  transmission for {keV} electrons, Nucl. Instr. and Meth. B 63~(3) (1992)
  345--358.
\newblock \href {http://dx.doi.org/10.1016/0168-583X(92)95119-C}
  {\path{doi:10.1016/0168-583X(92)95119-C}}.

\bibitem{KATRIN2005}
{KATRIN collaboration},
  \href{http://bibliothek.fzk.de/zb/berichte/FZKA7090.pdf}{{KATRIN} design
  report}, FZKA scientific report 7090.
\newline\urlprefix\url{http://bibliothek.fzk.de/zb/berichte/FZKA7090.pdf}

\bibitem{Gaisser2016}
T.~K. Gaisser, R.~Engel, E.~Resconi, {Cosmic Rays and Particle Physics},
  Cambridge University Press, 2016.

\bibitem{Bogdanova2006}
L.~N. Bogdanova, M.~G. Gavrilov, V.~N. Kornoukhov, A.~S. Starostin, {Cosmic
  muon flux at shallow depths underground}, Phys. Atom. Nucl. 69 (2006)
  1293--1298.
\newblock \href {http://dx.doi.org/10.1134/S1063778806080047}
  {\path{doi:10.1134/S1063778806080047}}.

\bibitem{Tanabashi2018}
M.~Tanabashi, et~al., {Review of Particle Physics}, Phys. Rev. D98~(3) (2018)
  030001.
\newblock \href {http://dx.doi.org/10.1103/PhysRevD.98.030001}
  {\path{doi:10.1103/PhysRevD.98.030001}}.

\bibitem{Formaggio2004}
J.~A. Formaggio, C.~J. Martoff, {Backgrounds to sensitive experiments
  underground}, Ann. Rev. Nucl. Part. Sci. 54 (2004) 361--412.
\newblock \href {http://dx.doi.org/10.1146/annurev.nucl.54.070103.181248}
  {\path{doi:10.1146/annurev.nucl.54.070103.181248}}.

\bibitem{Chatzidakis2015}
S.~Chatzidakis, S.~Chrysikopoulou, L.~Tsoukalas, {Developing a cosmic ray muon
  sampling capability for muon tomography and monitoring applications}, Nucl.
  Instr. and Meth. A 804 (2015) 33--42.
\newblock \href {http://dx.doi.org/10.1016/j.nima.2015.09.033}
  {\path{doi:10.1016/j.nima.2015.09.033}}.

\bibitem{Agostinelli2003}
S.~Agostinelli, et~al., {GEANT4: A Simulation toolkit}, Nucl. Instrum. Meth.
  A506 (2003) 250--303.
\newblock \href {http://dx.doi.org/10.1016/S0168-9002(03)01368-8}
  {\path{doi:10.1016/S0168-9002(03)01368-8}}.

\bibitem{Allison2006}
J.~Allison, et~al., {GEANT4 developments and applications}, IEEE Trans. Nucl.
  Sci. 53 (2006) 270.
\newblock \href {http://dx.doi.org/10.1109/TNS.2006.869826}
  {\path{doi:10.1109/TNS.2006.869826}}.

\bibitem{Allison2016}
J.~Allison, et~al., {Recent developments in GEANT4}, Nucl. Instrum. Meth. A 835
  (2016) 186--225.
\newblock \href {http://dx.doi.org/10.1016/j.nima.2016.06.125}
  {\path{doi:10.1016/j.nima.2016.06.125}}.

\bibitem{Groom2001}
D.~E. Groom, N.~V. Mokhov, S.~I. Striganov, {Muon stopping power and range
  tables 10-MeV to 100-TeV}, Atom. Data Nucl. Data Tabl. 78 (2001) 183--356.
\newblock \href {http://dx.doi.org/10.1006/adnd.2001.0861}
  {\path{doi:10.1006/adnd.2001.0861}}.

\bibitem{Bogdanov2006}
A.~G. Bogdanov, et~al., {Geant4 simulation of production and interaction of
  muons}, IEEE Trans. Nucl. Sci. 53 (2006) 513--519.
\newblock \href {http://dx.doi.org/10.1109/TNS.2006.872633}
  {\path{doi:10.1109/TNS.2006.872633}}.

\bibitem{PhDLeiber2014}
B.~Leiber,
  \href{https://publikationen.bibliothek.kit.edu/1000042415}{Investigations of
  background due to secondary electron emission in the {KATRIN}-experiment},
  Ph.D. thesis, Karlsruher Institut f{\"u}r Technologie (KIT) (2014).
\newline\urlprefix\url{https://publikationen.bibliothek.kit.edu/1000042415}

\bibitem{Henke1977}
B.~L. Henke, J.~A. Smith, D.~T. Attwood, {0.1--10-keV x-ray-induced electron
  emissions from solids---Models and secondary electron measurements}, J. Appl.
  Phys. 48~(5) (1977) 1852--1866.
\newblock \href {http://dx.doi.org/10.1063/1.323938}
  {\path{doi:10.1063/1.323938}}.

\bibitem{Chuklayaev1987}
S.~Chuklyaev, O.~Shchetinin, Emission of slow secondary electrons under the
  influence of ionizing radiation, Sov. J. At. Energy 63~(1) (1987) 560--562.
\newblock \href {http://dx.doi.org/10.1007/BF01125160}
  {\path{doi:10.1007/BF01125160}}.

\bibitem{Furman2002}
M.~A. Furman, M.~T.~F. Pivi, Probabilistic model for the simulation of
  secondary electron emission, Phys. Rev. ST Accel. Beams 5 (2002) 124404.
\newblock \href {http://dx.doi.org/10.1103/PhysRevSTAB.5.124404}
  {\path{doi:10.1103/PhysRevSTAB.5.124404}}.

\bibitem{PhDPrall2011}
M.~Prall,
  \href{https://www.researchgate.net/publication/238418212_Transmission_Function_of_the_Pre-Spectrometer_and_Systematic_Tests_of_the_Main-Spectrometer_Wire_Electrode_PHD_thesis}{Background
  reduction of the {KATRIN} spectrometers: Transmission function of the
  pre-spectrometer and systematic test of the main-spectrometer wire
  electrode}, Ph.D. thesis, Westf{\"a}lische Wilhelms-Universit{\"a}t
  M{\"u}nster (2011).
\newline\urlprefix\url{https://www.researchgate.net/publication/238418212_Transmission_Function_of_the_Pre-Spectrometer_and_Systematic_Tests_of_the_Main-Spectrometer_Wire_Electrode_PHD_thesis}

\bibitem{PhDSchwamm2004}
F.~Schwamm,
  \href{https://publikationen.bibliothek.kit.edu/1000000855}{Untergrunduntersuchungen
  f{\"u}r das {KATRIN}-experiment}, Ph.D. thesis, Universit{\"a}t Karlsruhe
  (2004).
\newline\urlprefix\url{https://publikationen.bibliothek.kit.edu/1000000855}

\bibitem{Amsbaugh2015}
J.~F. Amsbaugh, et~al., Focal-plane detector system for the {KATRIN}
  experiment, Nucl. Instr. and Meth. A 778~(0) (2015) 40--60.
\newblock \href {http://dx.doi.org/10.1016/j.nima.2014.12.116}
  {\path{doi:10.1016/j.nima.2014.12.116}}.

\bibitem{Arenz2018a}
M.~Arenz, et~al., {The KATRIN superconducting magnets: overview and first
  performance results}, JINST 13 (2018) T08005.
\newblock \href {http://dx.doi.org/10.1088/1748-0221/13/08/T08005}
  {\path{doi:10.1088/1748-0221/13/08/T08005}}.

\bibitem{MSVacuum2016}
M.~Arenz, et~al., Commissioning of the vacuum system of the {KATRIN} main
  spectrometer, JINST 11 (2016) P04011.
\newblock \href {http://dx.doi.org/10.1088/1748-0221/11/04/P04011}
  {\path{doi:10.1088/1748-0221/11/04/P04011}}.

\bibitem{Glueck2013}
F.~Gl{\"u}ck, et~al., Electromagnetic design of the large-volume air coil
  system of the {KATRIN} experiment, New J. Phys. 15~(8) (2013) 083025.
\newblock \href {http://dx.doi.org/10.1088/1367-2630/15/8/083025}
  {\path{doi:10.1088/1367-2630/15/8/083025}}.

\bibitem{Erhard2017}
M.~Erhard, et~al., {Technical design and commissioning of the KATRIN
  large-volume air coil system}, JINST 13~(02) (2018) P02003.
\newblock \href {http://dx.doi.org/10.1088/1748-0221/13/02/P02003}
  {\path{doi:10.1088/1748-0221/13/02/P02003}}.

\bibitem{Valerius2010}
K.~Valerius, {The wire electrode system for the KATRIN main spectrometer},
  Prog. Part. Nucl. Phys. 64 (2010) 291--293.
\newblock \href {http://dx.doi.org/10.1016/j.ppnp.2009.12.032}
  {\path{doi:10.1016/j.ppnp.2009.12.032}}.

\bibitem{Jordanov1994}
V.~T. Jordanov, G.~F. Knoll, {Digital synthesis of pulse shapes in real time
  for high resolution radiation spectroscopy}, Nucl. Instr. and Meth. A 345~(2)
  (1994) 337--345.
\newblock \href {http://dx.doi.org/10.1016/0168-9002(94)91011-1}
  {\path{doi:10.1016/0168-9002(94)91011-1}}.

\bibitem{how04}
M.~Howe, et~al., {Sudbury Neutrino Observatory} neutral current detector
  acquisition software overview, IEEE Trans. Nucl. Sci. 51~(3) (2004) 878--883,
  see also \url{http://orca.physics.unc.edu/}.
\newblock \href {http://dx.doi.org/10.1109/TNS.2004.829527}
  {\path{doi:10.1109/TNS.2004.829527}}.

\bibitem{Drexlin1998}
G.~Drexlin, {KARMEN upgrade and prospects at ESS}, Prog. Part. Nucl. Phys. 40
  (1998) 193--202.
\newblock \href {http://dx.doi.org/10.1016/S0146-6410(98)00025-8}
  {\path{doi:10.1016/S0146-6410(98)00025-8}}.

\bibitem{PhDWandkowsky2013}
N.~Wandkowsky, \href{http://nbn-resolving.org/urn:nbn:de:swb:90-366316}{Study
  of background and transmission properties of the {KATRIN} spectrometers},
  Ph.D. thesis, Karlsruher Institut f{\"u}r Technologie (KIT) (2013).
\newline\urlprefix\url{http://nbn-resolving.org/urn:nbn:de:swb:90-366316}

\bibitem{Furse2017}
D.~Furse, et~al., {Kassiopeia: A Modern, Extensible C++ Particle Tracking
  Package}, New J. Phys. 19~(5) (2017) 053012.
\newblock \href {http://dx.doi.org/10.1088/1367-2630/aa6950}
  {\path{doi:10.1088/1367-2630/aa6950}}.

\bibitem{Seiler1983}
H.~Seiler, Secondary electron emission in the scanning electron microscope, J.
  Appl. Phys. 54~(11) (1983) R1--R18.
\newblock \href {http://dx.doi.org/10.1063/1.332840}
  {\path{doi:10.1063/1.332840}}.

\bibitem{Greenwood2002}
J.~Greenwood, {The correct and incorrect generation of a cosine distribution of
  scattered particles for Monte-Carlo modelling of vacuum systems}, Vacuum
  67~(2) (2002) 217 -- 222.
\newblock \href {http://dx.doi.org/10.1016/S0042-207X(02)00173-2}
  {\path{doi:10.1016/S0042-207X(02)00173-2}}.

\bibitem{Chung1974}
M.~S. Chung, T.~E. Everhart, Simple calculation of energy distribution of
  low-energy secondary electrons emitted from metals under electron
  bombardment, J. Appl. Phys. 45~(2) (1974) 707--709.
\newblock \href {http://dx.doi.org/10.1063/1.1663306}
  {\path{doi:10.1063/1.1663306}}.

\bibitem{Joy2004}
D.~C. Joy, M.~S. Prasad, H.~M. Meyer, Experimental secondary electron spectra
  under {SEM} conditions, J. Microscopy 215~(1) (2004) 77--85.
\newblock \href {http://dx.doi.org/10.1111/j.0022-2720.2004.01345.x}
  {\path{doi:10.1111/j.0022-2720.2004.01345.x}}.

\bibitem{Behrens2017}
J.~Behrens, et~al., {A pulsed, mono-energetic and angular-selective UV
  photo-electron source for the commissioning of the KATRIN experiment}, Eur.
  Phys. J. C77~(6) (2017) 410.
\newblock \href {http://dx.doi.org/10.1140/epjc/s10052-017-4972-9}
  {\path{doi:10.1140/epjc/s10052-017-4972-9}}.

\bibitem{PhDBehrens2016}
J.~D. Behrens,
  \href{http://www.katrin.kit.edu/publikationen/phd_behrens.pdf}{{Design and
  commissioning of a mono-energetic photoelectron source and active background
  reduction by magnetic pulse at the KATRIN spectrometers}}, Ph.D. thesis,
  {Westf{\"a}lische Wilhelms-Universit{\"a}t M{\"u}nster} (2016).
\newline\urlprefix\url{http://www.katrin.kit.edu/publikationen/phd_behrens.pdf}

\bibitem{Abrahao2016}
T.~Abrah{\~a}o, et~al., {Cosmic-muon characterization and annual modulation
  measurement with Double Chooz detectors}, J. Cosmol. Astropart. Phys.
  1702~(02) (2017) 017.
\newblock \href {http://dx.doi.org/10.1088/1475-7516/2017/02/017}
  {\path{doi:10.1088/1475-7516/2017/02/017}}.

\bibitem{efron1994}
B.~Efron, R.~Tibshirani,
  \href{https://books.google.de/books?id=gLlpIUxRntoC}{{An Introduction to the
  Bootstrap}}, Chapman \& Hall/CRC Monographs on Statistics \& Applied
  Probability, Taylor \& Francis, 1994.
\newline\urlprefix\url{https://books.google.de/books?id=gLlpIUxRntoC}

\bibitem{PhDHarms2015}
F.~Harms,
  \href{https://publikationen.bibliothek.kit.edu/1000050027}{{Characterization
  and Minimization of Background Processes in the KATRIN Main Spectrometer}},
  Ph.D. thesis, Karlsruher Institut f{\"u}r Technologie (KIT) (2015).
\newline\urlprefix\url{https://publikationen.bibliothek.kit.edu/1000050027}

\bibitem{Feldman1997}
G.~J. Feldman, R.~D. Cousins, {A Unified approach to the classical statistical
  analysis of small signals}, Phys. Rev. D57 (1998) 3873--3889.
\newblock \href {http://dx.doi.org/10.1103/PhysRevD.57.3873}
  {\path{doi:10.1103/PhysRevD.57.3873}}.

\bibitem{GammaPaper}
{KATRIN Collaboration}, {Gamma-induced backgrounds in the KATRIN Main
  Spectrometer} (in preparation).

\bibitem{Drexlin2017}
G.~Drexlin, et~al., {Calculations and TPMC simulations of the reduction of
  radioactive decays of a noble gas by cryo-panels}, Vacuum 138~(Supplement C)
  (2017) 165--172.
\newblock \href {http://dx.doi.org/10.1016/j.vacuum.2016.12.013}
  {\path{doi:10.1016/j.vacuum.2016.12.013}}.

\bibitem{RadonPaper}
{KATRIN Collaboration}, {Radon-induced stored-particle background in the KATRIN
  Main Spectrometer} (in preparation).

\bibitem{Fraenkle2017}
F.~M. Fraenkle, {Background processes in the KATRIN main spectrometer}, J.
  Phys. Conf. Ser. 888~(1) (2017) 012070.
\newblock \href {http://dx.doi.org/10.1088/1742-6596/888/1/012070}
  {\path{doi:10.1088/1742-6596/888/1/012070}}.

\bibitem{arXivKleesiek2018}
M.~Kleesiek, et~al., {$\beta$-Decay Spectrum, Response Function and Statistical
  Model for Neutrino Mass Measurements with the {KATRIN} Experiment}\href
  {http://arxiv.org/abs/1806.00369} {\path{arXiv:1806.00369}}.

\end{thebibliography}

\end{document}